\documentclass[twocolumn]{aastex63}
\usepackage{lineno}
\usepackage{hyperref}
\usepackage[utf8]{inputenc}
\usepackage[reqno]{amsmath}
\usepackage{amsfonts}
\usepackage{enumerate}
\usepackage{ytableau}
\usepackage{graphicx}
\usepackage{indentfirst}
\usepackage[english]{babel}
\usepackage{overpic}
\usepackage[normal,footnotesize]{caption2}
\def\baselinestretch{1.25}
\usepackage{appendix}
\usepackage{amssymb}
\usepackage{amsthm}
\usepackage{epsfig,graphicx,epstopdf}
\usepackage{natbib}
\usepackage{multirow}
\usepackage{float}
\usepackage{threeparttable} 
\usepackage{booktabs}
\usepackage{manfnt}

\def\gray{$\gamma$-ray\ }
\def\grays{$\gamma$-rays\ }

\received{***}
\revised{***}
\accepted{***}
\submitjournal{ApJL}

\shorttitle{LHAASO Galaxy clusters}
\shortauthors{LHAASO Collaboration}

\begin{document}
\title{ Constraining the Cosmic-ray Energy Based on Observations of Nearby Galaxy Clusters by LHAASO }

\author{Zhen Cao}
\affiliation{Key Laboratory of Particle Astrophysics \& Experimental Physics Division \& Computing Center, Institute of High Energy Physics, Chinese Academy of Sciences, 100049 Beijing, China}
\affiliation{University of Chinese Academy of Sciences, 100049 Beijing, China}
\affiliation{TIANFU Cosmic Ray Research Center, Chengdu, Sichuan,  China}
 
\author{F. Aharonian}
\affiliation{TIANFU Cosmic Ray Research Center, Chengdu, Sichuan,  China}
\affiliation{University of Science and Technology of China, 230026 Hefei, Anhui, China}
\affiliation{Yerevan State University, 1 Alek Manukyan Street, Yerevan 0025, Armenia}
\affiliation{Max-Planck-Institut for Nuclear Physics, P.O. Box 103980, 69029  Heidelberg, Germany}
 
\author{Y.X. Bai}
\affiliation{Key Laboratory of Particle Astrophysics \& Experimental Physics Division \& Computing Center, Institute of High Energy Physics, Chinese Academy of Sciences, 100049 Beijing, China}
\affiliation{TIANFU Cosmic Ray Research Center, Chengdu, Sichuan,  China}
 
\author{Y.W. Bao}
\affiliation{Tsung-Dao Lee Institute \& School of Physics and Astronomy, Shanghai Jiao Tong University, 200240 Shanghai, China}
 
\author{D. Bastieri}
\affiliation{Center for Astrophysics, Guangzhou University, 510006 Guangzhou, Guangdong, China}
 
\author{X.J. Bi}
\affiliation{Key Laboratory of Particle Astrophysics \& Experimental Physics Division \& Computing Center, Institute of High Energy Physics, Chinese Academy of Sciences, 100049 Beijing, China}
\affiliation{University of Chinese Academy of Sciences, 100049 Beijing, China}
\affiliation{TIANFU Cosmic Ray Research Center, Chengdu, Sichuan,  China}
 
\author{Y.J. Bi}
\affiliation{Key Laboratory of Particle Astrophysics \& Experimental Physics Division \& Computing Center, Institute of High Energy Physics, Chinese Academy of Sciences, 100049 Beijing, China}
\affiliation{TIANFU Cosmic Ray Research Center, Chengdu, Sichuan,  China}
 
\author{W. Bian}
\affiliation{Tsung-Dao Lee Institute \& School of Physics and Astronomy, Shanghai Jiao Tong University, 200240 Shanghai, China}
 
\author{A.V. Bukevich}
\affiliation{Institute for Nuclear Research of Russian Academy of Sciences, 117312 Moscow, Russia}
 
\author{C.M. Cai}
\affiliation{School of Physical Science and Technology \&  School of Information Science and Technology, Southwest Jiaotong University, 610031 Chengdu, Sichuan, China}
 
% ...

\begin{abstract}

Galaxy clusters act as reservoirs of high-energy cosmic rays (CRs). As CRs propagate through the intracluster medium, they generate diffuse \grays detectable by arrays such as LHAASO. These \grays result from proton-proton ($pp$) collisions of very high-energy cosmic rays (VHECRs) or inverse Compton (IC) scattering of positron-electron pairs created by $p\gamma$ interactions of ultra-high-energy cosmic rays (UHECRs). We analyzed diffuse \gray emission from the Coma, Perseus, and Virgo clusters using LHAASO data. Diffuse emission was modeled as a disk of radius $R_{500}$ for each cluster while accounting for point sources. No significant diffuse emission was detected, yielding 95\% confidence level (C.L.) upper limits on the \gray flux: for WCDA (1–25 TeV) and KM2A ($>$25 TeV), less than (49.4, 13.7, 54.0) and (1.34, 1.14, 0.40) $\times 10^{-14}$ ph cm$^{-2}$ s$^{-1}$ for Coma, Perseus, and Virgo, respectively. The \gray upper limits can be used to derive model-independent constraints on the integral energy of CRp above 10 TeV (corresponding to the LHAASO observational range $>1\,\mathrm{TeV}$ under the $pp$ scenario) to be less than $(1.96, 0.59, 0.08) \times 10^{61}\,\mathrm{erg}$. The absence of detectable annuli/ring-like structures, indicative of cluster accretion or merging shocks, imposes further constraints on models in which the UHECRs are accelerated in the merging shocks of galaxy clusters.




  
\end{abstract}

\keywords{LHAASO, Galaxy clusters, \gray source, UHECRs, Coma, Perseus, Virgo}

\section{Introduction} \label{sec:intro}

Galaxy clusters, recognized as the largest gravitationally bound entities in the universe, serve as vital cosmological laboratories for investigating a wide range of astronomical phenomena. These clusters typically encompass hundreds to thousands of galaxies, all enveloped in hot, diffuse intracluster gas that extends across several megaparsecs. Radio observations have uncovered non-thermal radio halos and relics within these clusters, indicating the presence of a population of relativistic particles \citep{ferettiClustersGalaxiesObservational2012}.


Cosmic rays within galaxy clusters are generated and accelerated by multiple astrophysical sources. These include supernova-driven galactic winds, active galactic nuclei (AGNs) \citep{hintonRayEmissionAssociated2007}, large-scale intergalactic shocks arising from accretion and merger processes \citep{colafrancescoClustersGalaxiesDiffuse1998,ryuCosmologicalShockWaves2003,vannoniAccelerationRadiationUltrahigh2011,brunettiCosmicRaysGalaxy2014}, and turbulent intracluster magnetic fields \citep{brunettiAlfvenicReaccelerationRelativistic2005}. The magnetic fields in galaxy clusters, typically on the order of a few $\mu G$, effectively confine cosmic rays, allowing their accumulation \citep{vlkNonthermalEnergyContent1996,berezinskyClustersGalaxiesStorage1997a, govoniMAGNETICFIELDSCLUSTERS2004}. When these relativistic particles interact with the ambient matter, they can generate detectable high-energy $\gamma$-rays.


Studying galaxy clusters in the $\gamma$-ray band provides crucial insights into the efficiency of particle acceleration, magnetic confinement, and radiation processes involving cosmic rays. Several radiation mechanisms can give rise to $\gamma$-rays in galaxy clusters. The hadronic scenario, which involves inelastic collisions between cosmic ray protons (CRp) and thermal nuclei \citep{dennisonFormationRadioHalos1980}, is a major potential $\gamma$-ray radiation mechanism in galaxy clusters. Leptonic scenarios can also generate $\gamma$-rays. Relativistic CRp interact with target photon fields, such as the cosmic microwave background (CMB) and infrared background, producing high-energy $\gamma$-rays \citep{atoyanImplicationsNonthermalOrigin2000}. $\gamma$-ray emission can also arise from IC scattering of secondary electrons, which are produced by Bethe-Heitler processes involving relativistic protons and the CMB photon field, if UHECRs are accelerated within galaxy clusters, as predicted by some models \citep{kelnerEnergySpectraGamma2008}. Another mechanism involves IC radiation emitted by relativistic cosmic ray electrons (CRe), which originate from interactions between very high-energy (VHE) $\gamma$-rays and the diffuse extragalactic background radiation field.


Previous observations of $\gamma$-rays in the MeV to GeV energy range by EGRET \citep{reimerEGRETUpperLimits2003} provided only upper limits for several galaxy clusters. Similarly, the Whipple telescope reported upper limits at TeV energies for the Perseus and Abell 2029 clusters \citep{perkinsTeVGammaRay2006}. More recently, stereoscopic instruments such as H.E.S.S. and VERITAS have also set upper limits from TeV observations \citep{collaborationConstrainingCosmicrayPressure2023c, thehesscollaborationConstraintsMultiTeVParticle2009, arlenCONSTRAINTSCOSMICRAYS2012, perkinsTeVGammaRay2006}. The analysis of Fermi LAT data by \citet{xiDetectionGammarayEmission2018} revealed extended GeV emission from the direction of the Coma cluster, which was later confirmed by \citet{baghmanyanDetailedStudyExtended2022}.


Nearby galaxy clusters, owing to their large physical sizes, are anticipated to be extended sources of $\gamma$-rays. This characteristic makes large field-of-view instruments more effective for their study. In this context, LHAASO, with its $\sim$2.3 steradian FoV and unparalleled sensitivity in the VHE-UHE energy range, is highly suitable for such investigations.


In this Letter, we report the LHAASO observations of three nearby galaxy clusters: Coma, Perseus, and Virgo. Leveraging the $\gamma$-ray upper limits obtained from LHAASO observations, we impose stringent constraints on the energy budget of high-energy particles within these clusters above 10 TeV, corresponding to the $\gamma$-ray energy range observed by LHAASO ($>1\,\mathrm{TeV}$) under the $pp$ interaction scenario, in a model-independent manner. 


Such constraints can be extrapolated to lower energies to determine the total CRp energy budget of the galaxy cluster and, in turn, the ratio of CRp to thermal energy, \(X_{CRp}=\frac{E_{CRp}}{E_{th}}\). Here, \(E_{CRp}\) represents the integral energy of CRp above 1 GeV, and \(E_{th}\) denotes the total thermal energy. However, it must be noted that this extrapolation is highly dependent on the confinement of CRp within the galaxy cluster and is significantly model-dependent. This aspect will be elaborated on in detail in the subsequent text.

The structure of this paper is as follows. In Sec.\ref{sec2}, we present the fundamental background information on the selected clusters. In Sec.\ref{sec3}, we conduct an in-depth analysis of the LHAASO-WCDA and LHAASO-KM2A data. Finally, in Sec.\ref{sec:disc}, we discuss the derived cosmic-ray constraints and their astrophysical implications.

\section{Facts of the galaxy clusters}
\label{sec2}


Coma, Perseus, and Virgo are the three closest massive galaxy clusters, with redshifts of approximately 0.023 \citep{giovanniniHaloRadioSource1993}, 0.017 \citep{hitomicollaborationAtmosphericGasDynamics2018}, and 0.004 \citep{meiACSVirgoCluster2007}, and total masses $M_{500}$ of about $6.13\times10^{14} M_{\odot}$ \citep{planckcollaborationPlanckIntermediateResults2013}, $5.77\times10^{14} M_{\odot}$ \citep{consortiumProspectsGammaRay2023}, and $0.83\times10^{14} M_{\odot}$ \citep{simionescuWitnessingGrowthNearest2017}, respectively (as shown in Tab.\ref{tab:info}). Their proximity and large mass render these clusters ideal targets for studying cosmic rays and dark matter. 


Observations in other wavelengths have revealed that all three clusters exhibit radio halos, which are a clear indication of a substantial population of relativistic particles. Specifically, the Coma cluster possesses a giant radio halo, which is typically associated with cluster mergers \citep{cassanoConnectionGiantRadio2010}. This halo extends from the core of the Coma cluster and features diffuse radio relics near its virial radius \citep{thierbachDiffuseRadioEmission2003}. X-ray observations have shown that NGC 4839 is a subgroup in the south-west region of the Coma cluster, connected to its core by a faint X-ray bridge, suggesting a recent merger event \citep{churazovTempestuousLifeR5002021, malavasiSpiderItsWeb2020}. The presence of the giant radio halo, in tandem with the merger activities, implies that shock wave dissipation and turbulence resulting from the merging process play significant roles in injecting non-thermal energy into the system \citep{vannoniAccelerationRadiationUltrahigh2011}.


Perseus and Virgo are well-known as relaxed cool-core (CC) clusters \citep{petersonXraySpectroscopyCooling2006}, characterized by their dense cores. Perseus showcases a radio mini-halo (RMH) \citep{gendron-marsolaisHighresolutionVLALow2020, gendron-marsolaisVLAResolvesUnexpected2021} and X-ray cavities linked to the radio lobes of NGC 1275 (3C84) \citep{fabianChandraImagingComplex2000}. Similarly, the buoyant bubbles ascending around M87 \citep{churazovEvolutionBuoyantBubbles2001} suggest that active galactic nucleus (AGN) feedback plays a substantial role in Virgo. Moreover, the RMH in M87 is remarkable as one of the earliest discovered radio sources \citep{boltonVariableSourceRadio1948}. Effelsberg observations uncovered a faint radio halo around M86 in Virgo, indicating a past interaction between M86's intracluster medium (ICM) and low-density gas at the cluster's periphery \citep{vollmerDetectionRadioHalo2004}.


In these galaxy clusters, point sources contribute to the $\gamma$-ray emission. During the analysis, it is crucial to distinguish this emission from the extended $\gamma$-ray emission. Fermi observations indicate the presence of three high-energy point sources within the Coma galaxy cluster, which may be associated with radio galaxies or starburst galaxies \citep{baghmanyanDetailedStudyExtended2022}. Other studies have detected $\gamma$-ray emissions from NGC 1275 \citep{aleksicDetectionVeryhighEnergy2012} and IC 310 \citep{neronovVeryHighenergyRay2010} in the Perseus cluster, as well as from M87 \citep{aharonianGiantRadioGalaxy2003a} in the Virgo cluster. Significantly, several recent publications and alerts from LHAASO have reported detections of M87 \citep{collaborationDetectionVeryHighenergy2024}, NGC 1275 \citep{wangDetectionTwoTeV2024}, and IC 310 \citep{xiangLHAASODetectsRapid2024a, xiangLHAASODetectionRenewed2024}. Thus, we incorporate these sources as point sources in our analysis.

In recent years, multiple detectors have observed and placed constraints on the diffuse $\gamma$-ray emission from these clusters. HESS observations of the Coma cluster have constrained $X_{\text{CRp}}$ to values below 0.2, thereby ruling out the most favorable theoretical models \citep{thehesscollaborationConstraintsMultiTeVParticle2009}. Fermi observations have reported the detection of diffuse $\gamma$-ray emission from the Coma cluster, estimating $X_{\text{CRp}}$ to be around 1\% \citep{xiDetectionGammarayEmission2018}. Subsequent observations have confirmed this finding and provided further detailed analyses \citep{adamRayDetectionComa2021, baghmanyanDetailedStudyExtended2022}. However, these observations face challenges in differentiating between multiple point sources and extended sources, and the spectral indices obtained are relatively soft. Observations at other wavelengths, such as radio observations of the synchrotron emission from secondary pairs in the Coma cluster, yield constraints on \(X_{\text{CRp}}\) that range from less than 0.01\% to 28\%, depending on the assumptions made regarding the magnetic field. For the Perseus cluster, MAGIC telescope observations have constrained $X_{\text{CRp}}$ to less than approximately 1-2\% \citep{themagiccollaborationMAGICGammarayTelescope2010, magiccollaborationDeepObservationNGC2016, magiccollaborationConstrainingCosmicRays2012}. Preliminary simulations based on the expected performance of CTA indicate that the constraints on \(X_{\text{CRp}}\) could be an order of magnitude stronger than those from previous studies \citep{consortiumProspectsGammaRay2023}. HESS observations of the Virgo cluster have also constrained $X_{\text{CRp}}$ to values below 0.32 \citep{collaborationConstrainingCosmicrayPressure2023c}. These constraints are predicated on assumptions about the spatial and spectral distribution of cosmic rays and are applicable to different spatial scales.


In this paper, the constraints on $X_{\text{CRp}}$ are uniformly imposed within the $R_{500}$ range for each cluster. The coordinates of the cluster centers for Coma, Perseus, and Virgo are given as (\text{R.A.}, \text{Dec.}): (194.95$^{\circ}$, 27.98$^{\circ}$), (49.95$^{\circ}$, 41.51$^{\circ}$), and (187.70$^{\circ}$, 12.39$^{\circ}$), respectively. The Region of Interest (ROI) is defined as a circular area with a 3-degree radius, centered on the core positions of the three galaxy clusters. The $R_{500}$ values for these clusters, along with other physical parameters, are presented in Table \ref{tab:info}.


In the TeV energy range, $\gamma$-ray flux is attenuated due to pair production ($\gamma \gamma \rightarrow e^+e^-$) of $\gamma$-rays interacting with background photons from both the CMB and extragalactic background light (EBL). Appendix \ref{apxB} provides detailed information on EBL absorption within these clusters.

\section{Data analysis} \label{sec:intro}
\label{sec3}
\subsection{Analysis method}

\begin{table*}
\centering
\caption {Fitting Results of Coma, Perseus and Virgo cluster by WCDA and KM2A.}
\begin{tabular}{llllll}
\toprule[0.5mm]
&  & Spatial model & TS & TS$_{disk}$$^{a}$  & Flux U.L$^{b}$.\\ %
\hline
\multirow{5}{*}{Coma}   & \multirow{4}{*}{WCDA} & disk-2.3 & 6.25 & 6.25 & 61.4 \\
& & disk-2.7 & 7.33 & 7.33 & 69.0 \\
& & disk-2.3+1pt & 13.3 & 1.51 & 49.4 \\
& & disk-2.7+1pt & 14.5 & 2.75 & 57.1 \\
\cline{2-6}
& KM2A &  disk-3.0  & 3.51 & 3.51 & 1.34 \\

\hline
\multirow{3}{*}{Perseus} & \multirow{2}{*}{WCDA} & disk-2.3+2pt & 19.3 & 0.00 & 13.7\\
& & disk-2.7+2pt & 16.5 & 0.00 & 13.7\\
\cline{2-6}
& KM2A  &   disk-3.0 &1.08 &1.08 &1.14\\

\hline
\multirow{3}{*}{Virgo}   & \multirow{2}{*}{WCDA} & disk-2.3+1pt & 31.72 & 0.43 & 54.0\\
& & disk-2.7+1pt & 31.47 & 0.19 & 44.9  \\
\cline{2-6}
& KM2A &   disk-3.0  &0.53 &0.53&0.40\\
\bottomrule[0.4mm]
\end{tabular}
\\
{\footnotesize 
$^a$ The increase in the TS value resulting from the inclusion of the disk model \\ under the assumption of the existence of other sources in the Spatial model. \\

$^b$ 95\% C.L. Flux U.L. of the disk component. (For WCDA, the integration energy range is 1-25 TeV, \\
while for KM2A, it is above 25 TeV. The unit is 10$^{-14}$ photon cm$^{-2}$ s$^{-1}$.)
}
\label{tab:fits}
\end{table*}


LHAASO represents a pioneering, ground-based $\gamma$-ray and cosmic-ray detector array of the new generation. With its broad bandwidth, wide field of view, and all-weather adaptability, it enables comprehensive detection of extended emissions, including those from galaxy clusters.

LHAASO consists of two detectors for $\gamma$-ray detection. The lower-energy data ($E < 25 \, \text{TeV}$) are collected by the Water Cherenkov Detector Array (WCDA), which covers an area of $78,000 \, \text{m}^2$. The data collection period spans from March 5, 2021, to July 31, 2023, with a total effective observation time of 784 days. The high-energy $\gamma$-ray data ($E > 10 \, \text{TeV}$) are recorded by the Kilometer Square Array (KM2A), which consists of 5195 electromagnetic particle detectors (EDs) and 1188 muon detectors (MDs). The KM2A data incorporate observations from the half-array phase, through the three-quarter-array phase, to the full-array phase, covering the period from December 27, 2019, to January 31, 2024, with a total effective observation time of 1389.5 days.

After data selection, the number of events used in this analysis is $4.69 \times 10^9$ for WCDA and $1.90 \times 10^9$ for KM2A. For more details regarding the array and the reconstruction of WCDA and KM2A, refer to \citep{aharonianObservationCrabNebula2021a, aharonianPerformanceLHAASOWCDAObservation2021b}.

\begin{figure*}
    \centering
     \includegraphics[width=0.3\linewidth]{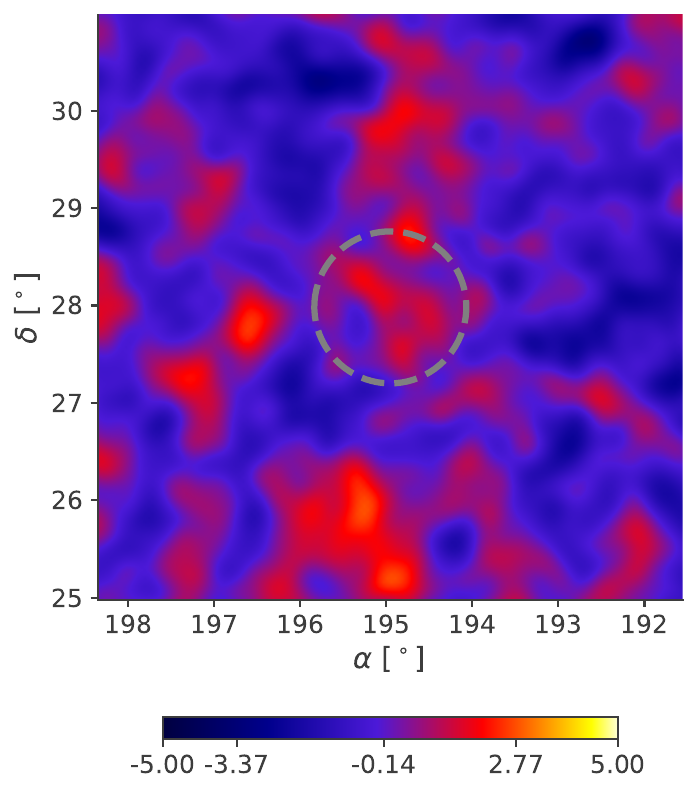}
    \includegraphics[width=0.302\linewidth]{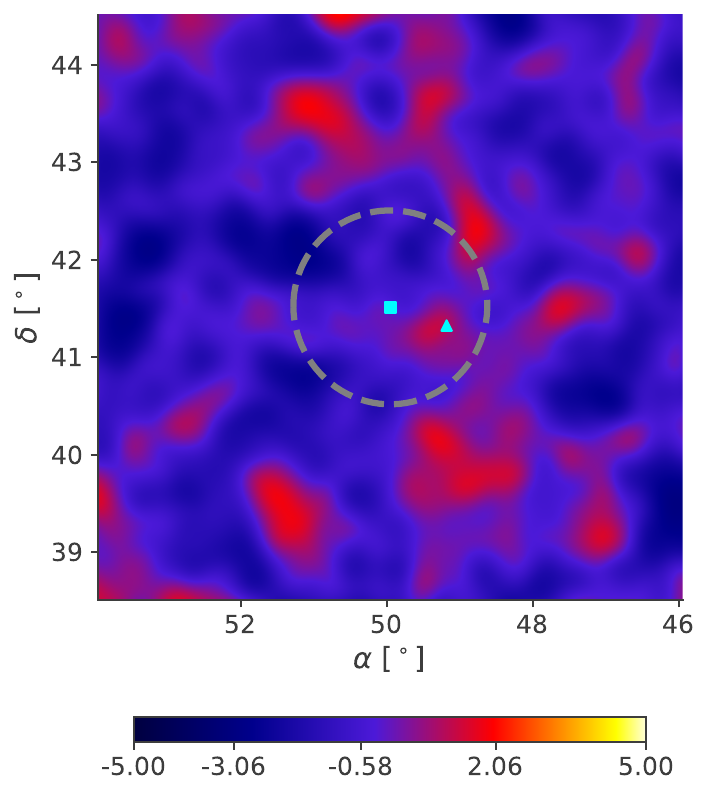}
    \includegraphics[width=0.3\linewidth]{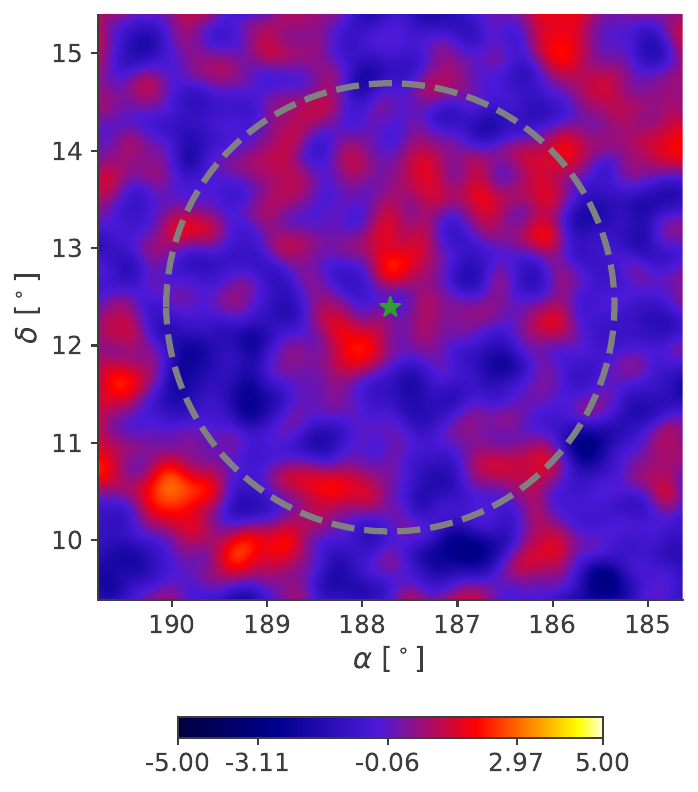}

    \includegraphics[width=0.3\linewidth]{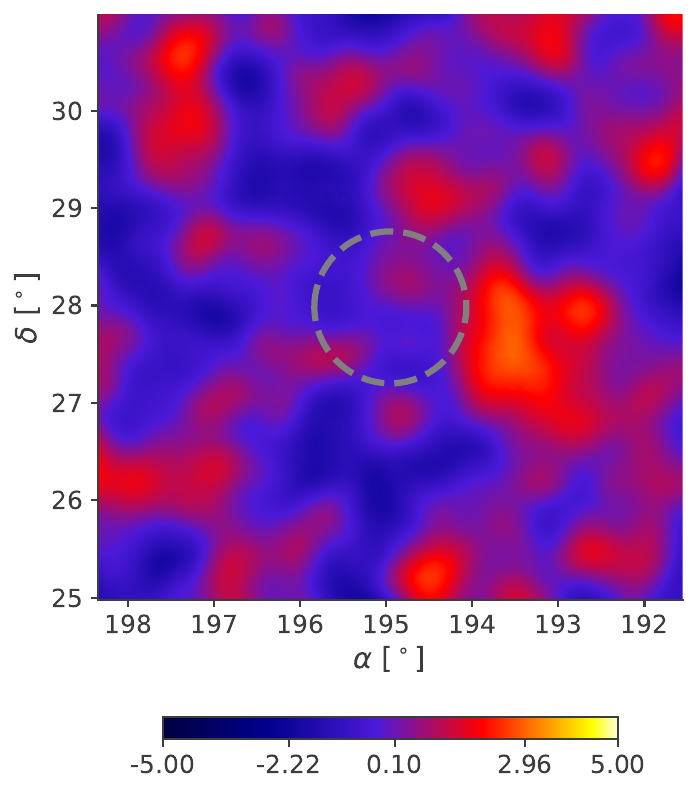}
    \includegraphics[width=0.302\linewidth]{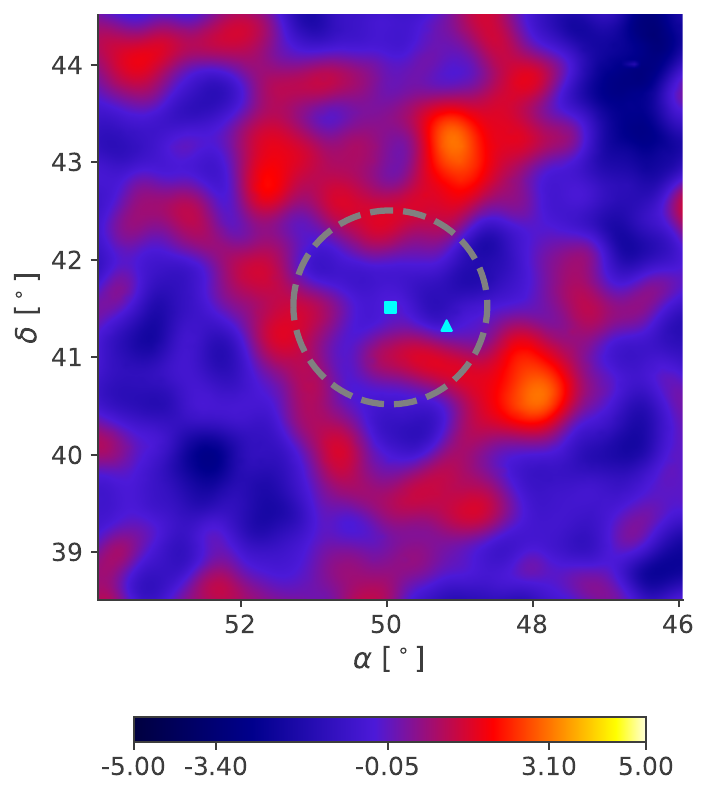}
    \includegraphics[width=0.3\linewidth]{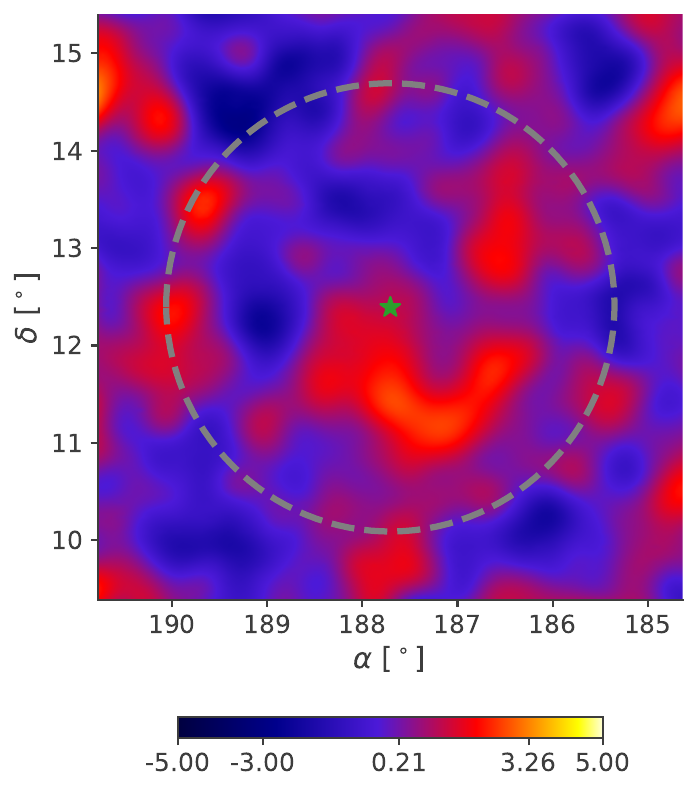}  
    \caption{The upper panel shows the WCDA skymaps (1-25 TeV) with the contribution of known point sources subtracted, while the lower panel displays the KM2A skymaps ($>$ 25 TeV). From left to right, they correspond to the Coma, Perseus, and Virgo clusters. The $R_{500}$ radius of these clusters is respectively indicated by gray dashed lines. The known \grays point sources NGC 1275 and IC 310 in the Perseus cluster are respectively marked with cyan squares and upward triangles, while M87 in the Virgo cluster is indicated by a green star.}
    \label{fig:fig1}
\end{figure*}


The WCDA leverages the number of triggered PMT units per event, denoted as $N_{\text{hit}}$, to approximate the event energy. All events are partitioned into six $N_{\text{hit}}$ bins: 60 - 100, 100 - 200, 200 - 300, 300 - 500, 500 - 800, and $>$ 800. 

For the KM2A, its datasets are divided into two energy bins per decade, each with a bin width of $\Delta\log_{10}(E) = 0.2$, based on the reconstructed energy. Given EBL absorption and uncertainties in EBL models, the use of higher-energy bins becomes unnecessary. 

Based on the reconstructed direction of each event, all events are binned into skymaps with a pixel size of $0.1^{\circ} \times 0.1^{\circ}$. The cosmic-ray background is estimated using the direct integration method \citep{fleysherTestsStatisticalSignificance2004b}.


The three-dimensional (3D) algorithm was employed to simultaneously fit both the spectrum and morphology of the source. In this study, we compute the source significance using ${\rm TS} = 2\log\left({\frac{\mathcal{L}_{1}}{\mathcal{L}_{0}}}\right)$. Here, $\mathcal{L}_{1}$ represents the maximum likelihood under the alternative hypothesis we intend to test, while $\mathcal{L}_{0}$ denotes the maximum likelihood under the null hypothesis. According to Wilks' Theorem \citep{wilksLargeSampleDistributionLikelihood1938a}, the TS value follows a chi-squared distribution with degrees of freedom equal to the difference between those of the null and alternative hypotheses.

To determine the significance in the sky map for this work, we assume that the source is a point source. For WCDA in the energy range of 1-25 TeV, we adopt a power-law spectrum with an index of 2.7; for KM2A at energies above 25 TeV, we use an index of 3.5. In each pixel, the flux is the only free parameter. In line with Wilks' Theorem, we consider $\pm \sqrt{\rm TS}$ as the measure of significance.


To distinguish between the point-source and extended components within these regions, this study employs a likelihood-fitting approach. For the extended radiation, we describe it using a disk centered at the core of each cluster, with a radius equal to its respective $R_{500}$ value. Regarding the energy spectrum of the disk model, we adopt a power-law model with EBL absorption, as presented in the following equation:
\begin{equation} \label{Equ: spec}
    f(E)=A\cdot\left(\frac{E}{E_{0}}\right)^{\Gamma}\cdot e^{-\tau_{ebl}(E,z)}
\end{equation}
Here, we fix the intrinsic spectral index of the extended radiation. The uncertainty in the spectral index of the extended radiation gives rise to systematic errors. To address this, we test spectral index assumptions of -2.3 and -2.7. Additionally, since LHAASO has reported detections of M87, NGC 1275, and IC 310, we incorporate these sources as point sources in our analysis, using the same spectral model assumption as described in Equation \ref{Equ: spec}. For the hotspot within the Coma cluster, we evaluate models both with and without a point-source component to assess systematic uncertainties, fixing the intrinsic index to -2.3.

\subsection{Results}

\begin{figure*}
    \centering
     \includegraphics[width=0.47\linewidth]{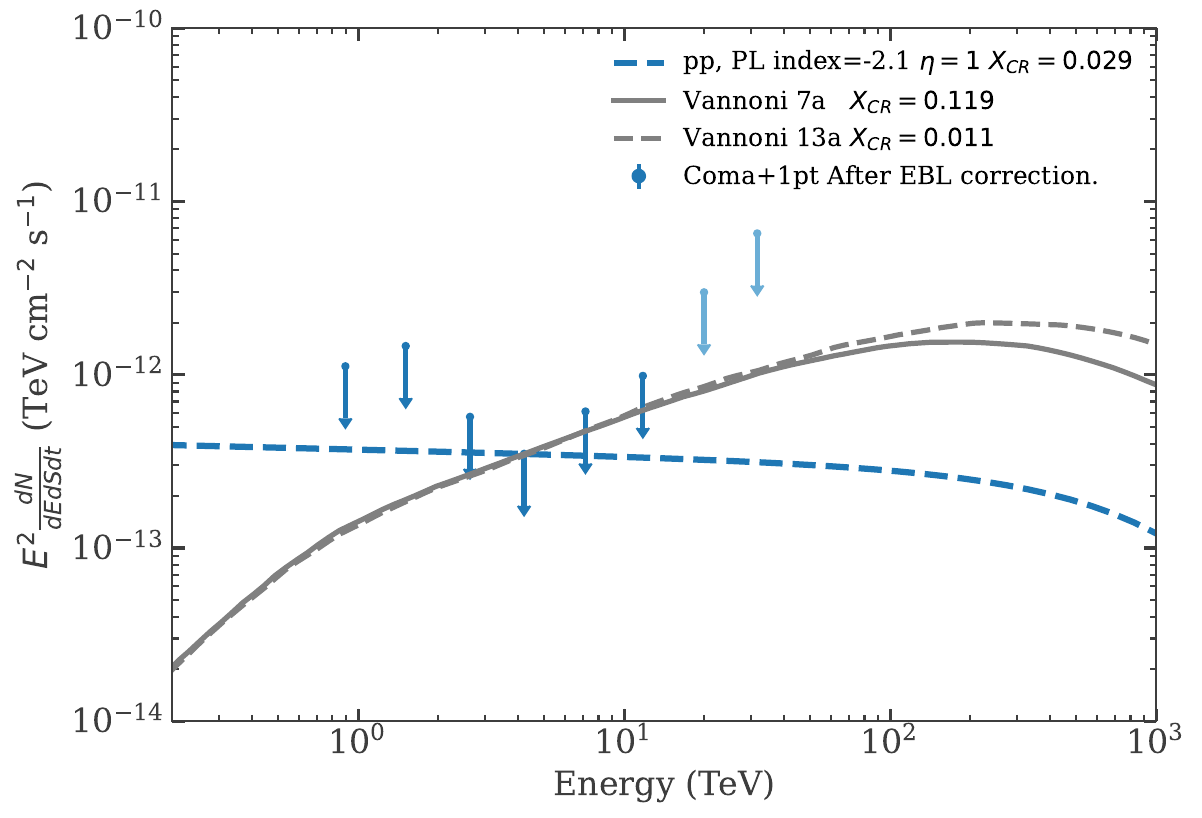}
     \includegraphics[width=0.45\linewidth]{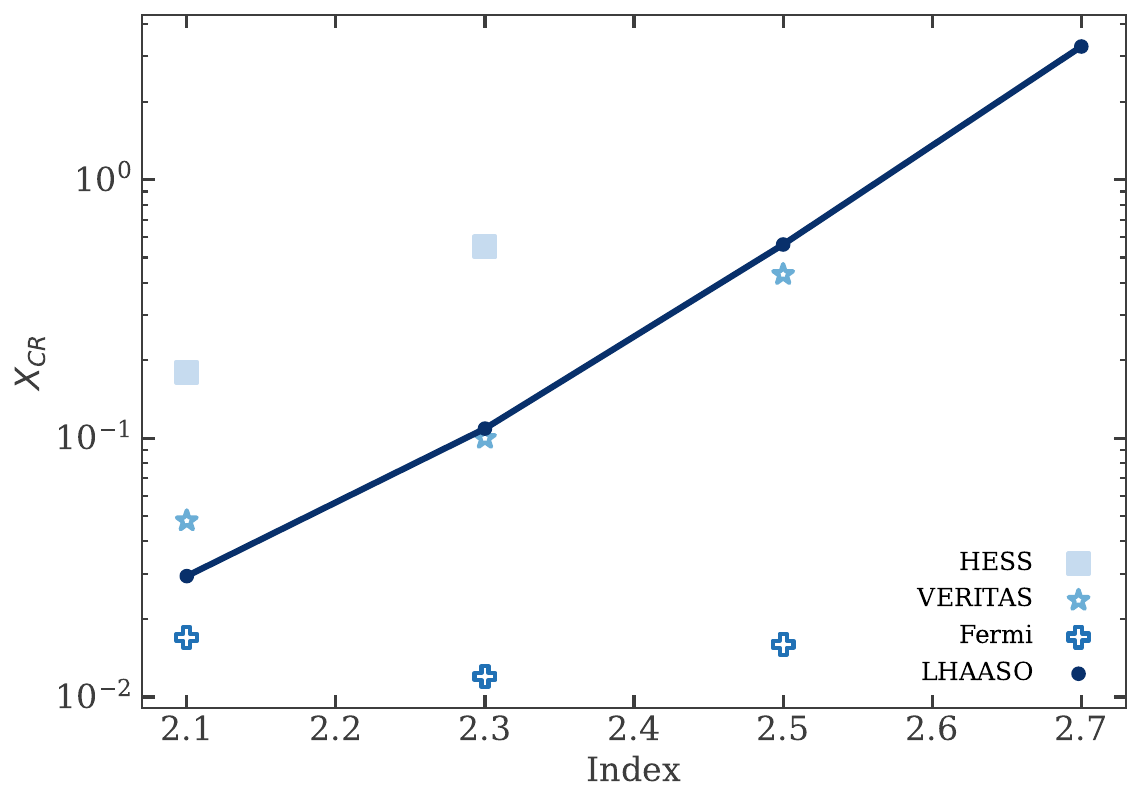}

    \includegraphics[width=0.47\linewidth]{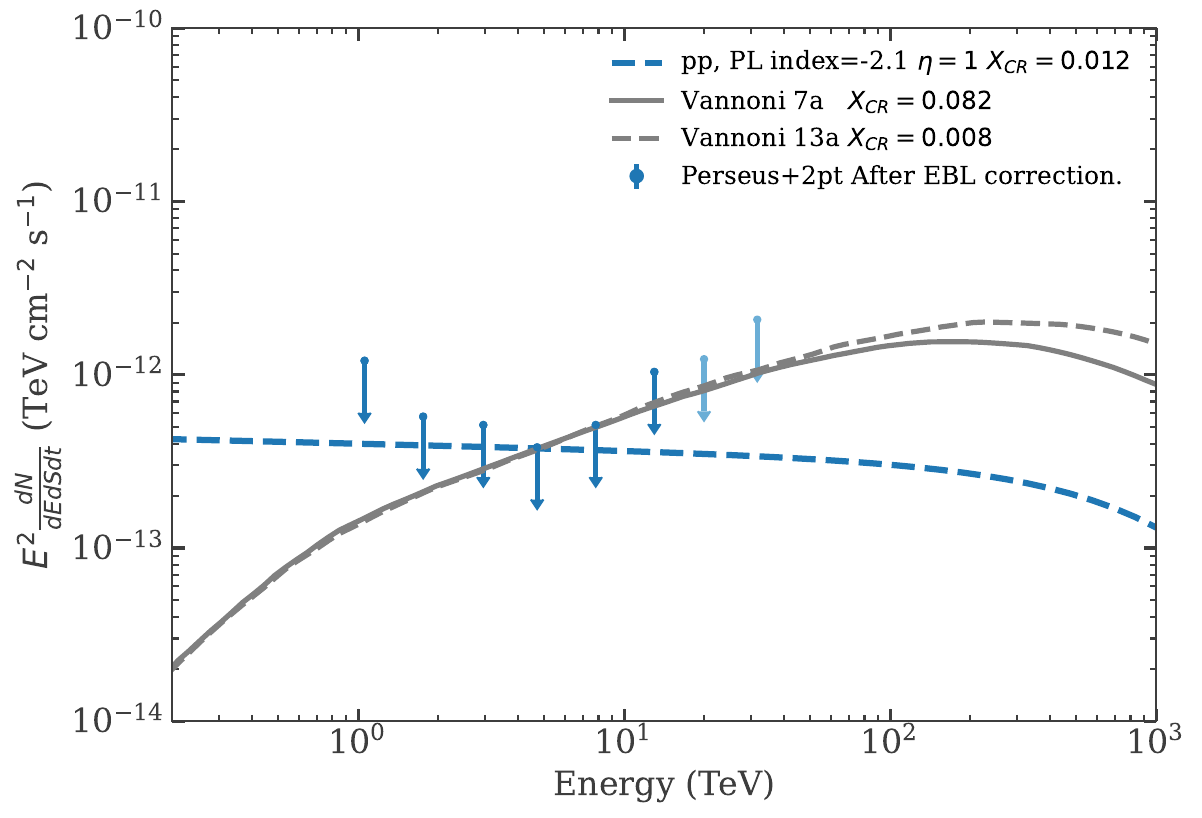}
     \includegraphics[width=0.45\linewidth]{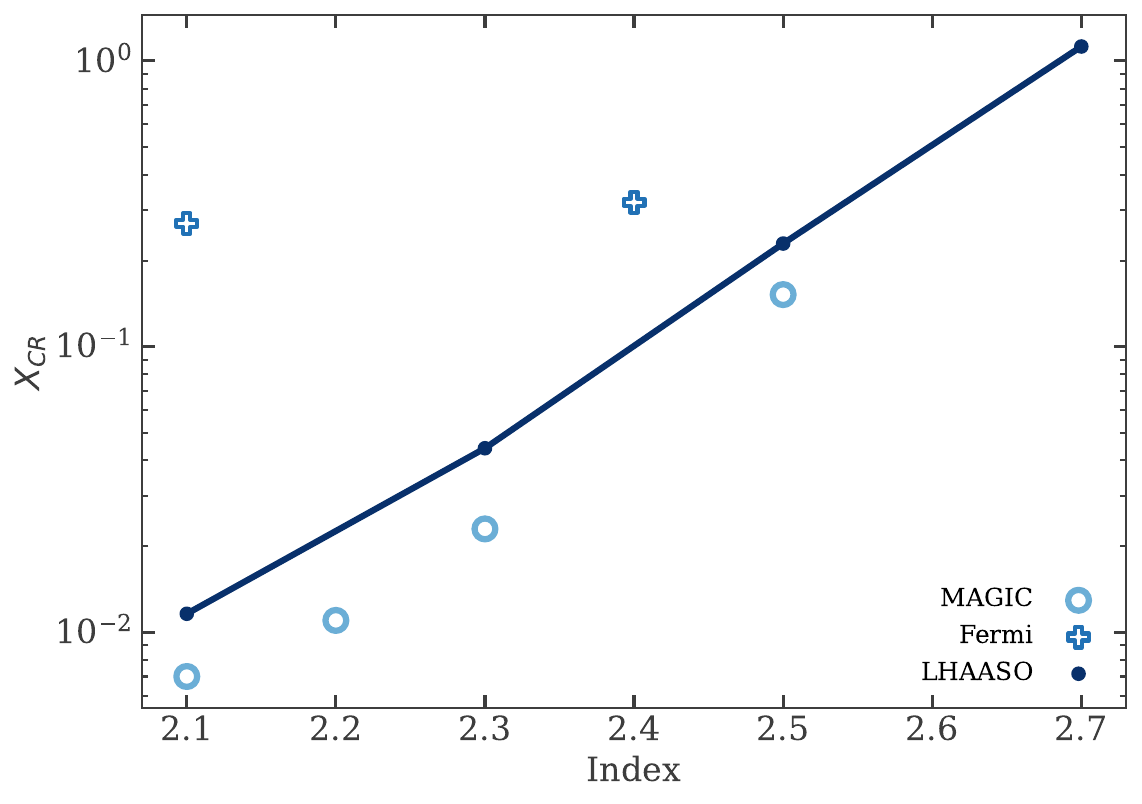}
    
    \includegraphics[width=0.47\linewidth]{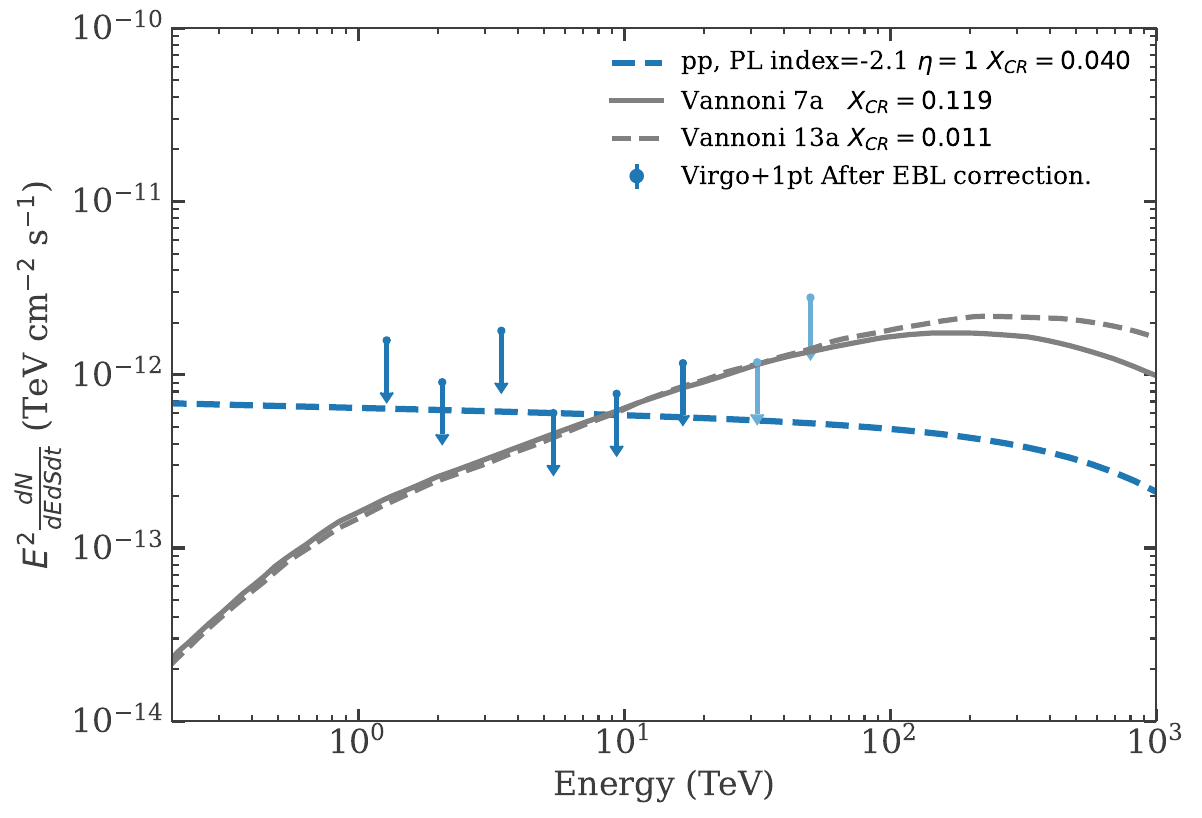}
     \includegraphics[width=0.45\linewidth]{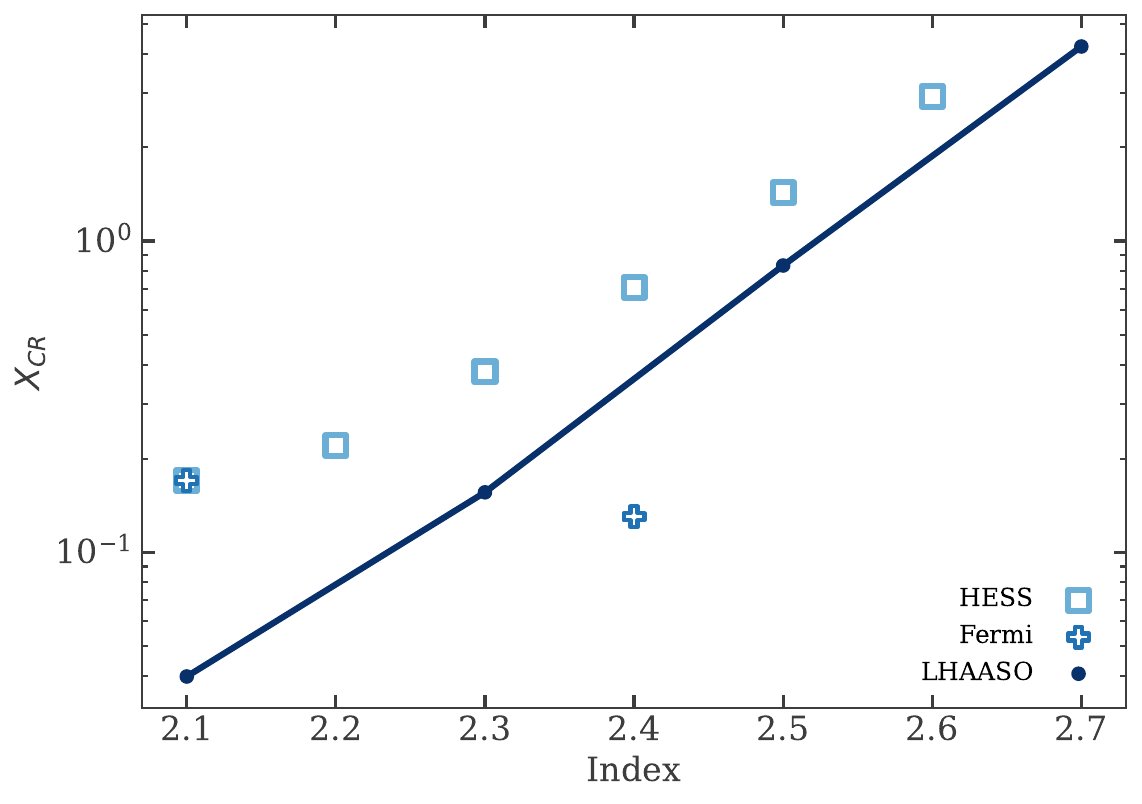}
    
    \caption{Left panel: The 95\% C.L. flux upper limits for the extended emission within the $R_{500}$ radius of the three galaxy clusters, after correction for the EBL, are presented. The darker-blue and lighter-blue arrows represent the data from WCDA and KM2A, respectively. The gray solid and dashed lines represent the $\gamma$-ray fluxes scaled to the strongest constraint for the IC process of secondary electrons produced by the $p-\gamma$ interactions of accelerated cosmic rays with linear and nonlinear DSA scenarios, respectively, as shown in Figs. 7 and 13 of \citet{vannoniAccelerationRadiationUltrahigh2011}. The blue dashed line represents the $\gamma$-ray flux scaled to the strongest constraint for hadronic processes. From top to bottom, the corresponding clusters are the Coma, Perseus, and Virgo clusters. Right panel: The 95\% C.L. upper-limit constraints on $X_{\text{CRp}}$ for these three galaxy clusters, under the hypothesis of different CRp spectral indices, are shown. The sequence of galaxy clusters from top to bottom is consistent with the order presented in the left panel. The solid squares in the first plot on the right panel represent the strongest constraints (99\% C.L.) previously obtained by HESS \citep{thehesscollaborationConstraintsMultiTeVParticle2009}. The hollow pentagrams and crosses represent VERITAS's and Fermi's strongest constraints (99\% C.L.) on $X_{\text{CRp}}$ within the virial radius of Coma \citep{arlenCONSTRAINTSCOSMICRAYS2012}. The hollow circles in the second plot on the right denote the 95\% C.L. constraints on $X_{\text{CRp}}$ within the virial radius of the Perseus cluster, as determined by MAGIC \citep{magiccollaborationDeepObservationNGC2016}. The hollow squares in the bottom-most plot represent the 99.7\% C.L. constraints on $X_{\text{CRp}}$ within the Virgo cluster at $\sim 20 \, \text{kpc}$, as determined by HESS \citep{collaborationConstrainingCosmicrayPressure2023c}. In the bottom two panels, the open crosses represent the 95\% C.L. upper limits provided by Fermi \citep{ackermannGeVGAMMARAYFLUX2010}.
}

    \label{fig:fig2}
\end{figure*}


The fitting results are presented in Table \ref{tab:fits}. The significance maps of the Coma, Perseus, and Virgo clusters, with all point sources subtracted, for the WCDA and KM2A are shown in Figure \ref{fig:fig1}. Under all model assumptions, we did not detect any significant extended radiation in these clusters. Consequently, we obtained the 95\% C.L. upper limits on the extended radiation of these clusters after EBL correction, using likelihood scans, as shown in the left panel of Figure \ref{fig:fig2}. The darker blue and lighter blue represent the upper limits obtained by WCDA and KM2A, respectively. 

For the WCDA data, the consideration of point sources has an impact of approximately 2\% on the strongest-constraint upper limit of the Coma cluster, which plays a dominant role in the constraints on $X_{\text{CRp}}$. The assumptions of different spectral indices introduce impacts of about 7\%, 4\%, and 10\% on the strongest-constraint upper limits for the Coma, Perseus, and Virgo clusters, respectively. 

Due to the negligible impact of the two aforementioned uncertainties on the constraints, subsequent analyses will uniformly adopt a spectral-index assumption of $\Gamma = -2.3$ and will invariably consider point sources to derive the upper limits of the disk component for these galaxy clusters. For the KM2A, since there are no significant point sources or hotspots in the ROI, we only consider a disk component with a fixed spectral index of $\Gamma = -3.0$.

\section{Discussion and Conclusion} \label{sec:disc}

Based on the aforementioned constraints on $\gamma$-rays, we can place limits on the energy budget of CRp in galaxy clusters. LHAASO measured the \gray emissions above $1~\rm TeV$, which correspond to the CRp energy of more than $10~\rm TeV$ \citep{kafexhiuParametrizationGammarayProduction2014a, kelnerEnergySpectraGamma2006d}. Thus, the \gray upper limits we derived above can be used to derive the model-independent CRp energy budget in this energy range directly, which is listed in Table \ref{tab:xcr}.

The CRp are accelerated by various shocks and turbulent environments within galaxy clusters and then diffuse throughout the cluster. The large scale of galaxy clusters and their magnetic fields (on the order of a few $\mu$G) endows CRp with an escape timescale comparable to cosmological timescales. Furthermore, due to the low density of the ICM, the cooling timescale for CRp exceeds the Hubble time, signifying that CRp in galaxy clusters retains information about their acceleration and propagation. The total CRp energy in the cluster reflects the efficiency of cosmic-ray acceleration. Therefore, if we assume that the extended $\gamma$-ray radiation from galaxy clusters is due to the collision of CRp with the ICM, the upper limits previously derived can constrain the physics related to CRp acceleration and propagation for these three galaxy clusters. 


In scenarios where CRs accelerated within the cluster experience full confinement, the resulting CRp population would continuously accumulate over time rather than reaching a stationary state. Under the assumption of energy-independent hadronic interaction cross sections, the resulting $\gamma$-ray emission spectrum would then reflect this time-integrated CRp population. The energy budget integrated over the entire energy spectrum of CRp can be used to derive $X_{CR}$. However, the confinement of CRs is determined by the properties of turbulence in these regions, which can result in an energy-dependent confinement time \citep{blasiGAMMARAYSCLUSTERS2007}. The confinement can then be effective only below some critical energy $E_c$. Below $E_c$, the confinement time is larger than the Hubble time, and the \gray spectrum is the same as the injected/accelerated one. For energies above $E_c$, the \gray spectrum is softened due to the escape of higher-energy CRs. In LHAASO observations, the most stringent constraints come from the first several energy bins of WCDA of several $\rm TeV$, which correspond to the CRp energy of dozens of TeV. Thus, if $E_c > 30~\rm TeV$, the constraints derived from LHAASO can be extrapolated to lower energies according to the acceleration spectrum, which is typically a power law. On the other hand, if $E_c < 30~\rm TeV$, we expect a softening of the \gray spectrum in the WCDA energy band. Thus, the CRp energy budget above $10~\rm TeV$ we derived here cannot be extrapolated to lower energies to constrain $X_{CR}$. 

In addition to the uncertain confinement conditions, other significant uncertainties persist, particularly concerning the energy and spatial distributions of CRp and the ICM. Under the assumptions that the CRp energy spectrum follows a power-law distribution and that cosmic rays blow 30 TeV can be completely confined within galaxy clusters, when considering the Diffusive Shock Acceleration (DSA) of CRp, we expect the spatially integrated CRp energy spectrum to be the same as the CRp injection spectrum and to be hard. For example, it could have an index of -2.1 or -2.3. However, a low shock Mach number \citep{gabiciNonthermalRadiationClusters2003} and propagation effects may lead to a softer spectral index in the local spatial region, such as near the CRp source. Thus, we investigated four scenarios with spectral indices of -2.1, -2.3, -2.5, and -2.7. It is important to note that both theoretical analyses and simulation results generally support a hard spatially integrated CRp energy spectrum (index $>$ -2.5) \citep{vannoniAccelerationRadiationUltrahigh2011, pinzkeSimulatingGrayEmission2010}. Therefore, the constraints based on an index of -2.1 or -2.3 are likely more realistic and deserve emphasis.


%

Here, we assume a spherically symmetric distribution for both CRp and the ICM. Hypothesizing that the CRp density distribution adheres to the same form as the ICM density distribution, this is to maintain consistency with similar previous studies \citep{thehesscollaborationConstraintsMultiTeVParticle2009, arlenCONSTRAINTSCOSMICRAYS2012, magiccollaborationDeepObservationNGC2016, collaborationConstrainingCosmicrayPressure2023c, ackermannGeVGAMMARAYFLUX2010, consortiumProspectsGammaRay2023}. If we further assume that all CRp are confined within the galaxy clusters, we can derive $X_{CR}$ and list the values in Tab.\ref{tab:xcr}. However, it must be noted that the $X_{CR}$ we derived here is highly model-dependent.

This results are also shown in Figure \ref{fig:fig2}. The blue dashed line in the left panel  represents the expected $\gamma$-ray flux. This flux is computed using the parametrization of the $\gamma$-ray production cross-section in $pp$ collisions, as derived in \citep{kafexhiuParametrizationGammarayProduction2014a}. The calculation is carried out under the most stringent constraints, assuming a spectral index of $-2.1$. Under this assumption, the Coma, Perseus, and Virgo clusters constrain the value of $X_{\text{CRp}}$ to be less than approximately $0.029$, $0.012$, and $0.04$, respectively. The total thermal energy within the $R_{500}$ radius of the three galaxy clusters, denoted as $E_{th_{500}}$, can be found in Table \ref{tab:info}. Constraints on $X_{CRp}$ under other assumptions of spectral index are presented in Table \ref{tab:xcr}.

Radio observations of the synchrotron radiation of electrons in galaxy clusters have indirectly constrained the energy density of relativistic protons in the Coma galaxy cluster. Assuming a magnetic field of $2\,\mu$G in the galaxy cluster, the constraints on \(X_{\text{CRp}}\) are \(X_{\text{CRp}} < 0.009\%\), \(0.01\%\), and \(0.07\%\) based on cosmic-ray index assumptions of 2.1, 2.3, and 2.5, respectively \citep{reimerComaClusterRay2004}. Observations of the magnetic field in Coma cluster indicate that the magnetic field at the center of the it ranges from $3.9$ to $5.4\,\mu$G \citep{bonafedeComaClusterMagnetic2010}. This implies that the constraints from radio observations are at least an order of magnitude tighter than $\gamma$-ray constraints, highlighting the complementarity of multi-wavelength approaches. However, our $\gamma$-ray constraints remain complementary, particularly for scenarios with suppressed magnetic fields or non-standard secondary pair cooling \citep{ensslinModificationClusterRadio2002, brunettiSunyaevZeldovichEffectResponsible2013}. For comparison, the constraints on \(X_{cr}\) from other $\gamma$-ray detectors are presented on the right panel of Figure \ref{fig:fig2}.

\begin{table*}
\centering
\caption {Constraints on the $X_{CR}$ in these three clusters for different assumptions and scenario.}
\begin{tabular}{llllll}
\toprule[0.6mm]
& Scenario  & Assumptions$^{a}$ & $X_{CR}^{b}$ \quad\quad\quad & $E_{CRp_{500}}^{c}$ & $E_{CRp_{500}}^{d}(>10 \text{TeV})$ \\ %

\toprule[0.3mm]

\multirow{6}{*}{Coma\quad\quad\quad}   & \multirow{4}{*}{pp} & index=-2.1 \quad\quad\quad\quad\quad\quad & 0.029 \quad & $7.77 \times 10^{61} $ & $1.96 \times 10^{61} $ \\
& & index=-2.3 \quad\quad\quad\quad\quad\quad & 0.10\quad & $ 2.89 \times 10^{62} $ & $1.63 \times 10^{61} $ \\
& & index=-2.5 \quad\quad\quad\quad\quad\quad & 0.55 \quad & $ 1.49 \times 10^{63} $ & $1.46 \times 10^{61} $ \\
& & index=-2.7  \quad\quad\quad\quad\quad\quad & 3.22 \quad & $ 8.66 \times 10^{63} $ & $1.39 \times 10^{61} $ \\
\cline{2-6}
& \multirow{2}{*}{$p\gamma$+IC} & linear DSA \quad $\xi=4$& 0.12 & $ 3.23 \times 10^{62} $ & N/A \\
& & nonlinear DSA $\xi=7$& 0.011 & $ 2.96 \times 10^{61}$ & N/A \\


\hline
\multirow{6}{*}{Perseus\quad\quad\quad}   & \multirow{4}{*}{pp} & index=-2.1  & 0.012 & $2.48 \times 10^{61} $ & $ 5.98 \times 10^{60} $ \\
& & index=-2.3  & 0.045 & $9.42 \times 10^{61} $ & $ 5.09 \times 10^{60} $ \\
& & index=-2.5  & 0.24 & $4.90 \times 10^{62} $ & $ 4.62 \times 10^{60} $ \\
& & index=-2.7  & 1.16 & $2.40 \times 10^{63} $ & $ 3.67 \times 10^{60} $ \\
\cline{2-6}
& \multirow{2}{*}{$p\gamma$+IC} & linear DSA \quad $\xi=4$& 0.082 & $1.70 \times 10^{62} $ & N/A \\
& & nonlinear DSA $\xi=7$& 0.008 & $1.66 \times 10^{61} $ & N/A \\

\hline
\multirow{6}{*}{Virgo\quad\quad\quad}   & \multirow{4}{*}{pp} & index=-2.1  & 0.038 & $3.21 \times 10^{60} $ & $ 8.33 \times 10^{59} $ \\
& & index=-2.3  & 0.15 & $1.25 \times 10^{61} $ & $ 7.30 \times 10^{59} $ \\
& & index=-2.5  & 0.80 & $6.72 \times 10^{61} $ & $ 6.81 \times 10^{59} $ \\
& & index=-2.7  & 4.03 & $3.40 \times 10^{62} $ & $ 5.59 \times 10^{59} $ \\
\cline{2-6}
& \multirow{2}{*}{$p\gamma$+IC} & linear DSA \quad $\xi=4$& 0.12 & $1.00 \times 10^{61} $ & N/A \\
& & nonlinear DSA $\xi=7$& 0.011 & $9.27 \times 10^{59} $ & N/A \\

\bottomrule[0.4mm]
\end{tabular}
\\
{\footnotesize 
$^a$ For the $ p\gamma + \text{IC} $ process, here, $ \xi $ represents the compression factor \\ of the shock wave, assuming a shock velocity of 2000 km/s and an upstream magnetic field of 0.3 $\mu G$. \\ For further details, please refer to the article \citet{vannoniAccelerationRadiationUltrahigh2011}. \\
$^b$ 95\% C.L. upper limit for $X_{\text{CR}}$. \\
$^c$ 95\% C.L. upper limit for the total energy of CRp within the $R_{500}$ radius. The unit is erg. \\
$^d$ 95\% C.L. upper limit for the total energy of CRp with energy $>$ 10 TeV within the $R_{500}$ radius. The unit is erg. \\}
\label{tab:xcr}
\end{table*}

In addition to the $pp$ interaction of CR protons with ambient gas, other mechanisms in galaxy clusters, such as the IC processes of relativistic electrons or the $p\gamma$ processes of UHECR accelerated by accretion/merging shocks, can also generate $\gamma$-ray radiation. Indeed, galaxy clusters are believed to be the most ideal sites for the acceleration of UHECRs, given their large size and thus long-enough confinement time \citep{hillasOriginUltraHighEnergyCosmic1984, vannoniAccelerationRadiationUltrahigh2011, zirakashviliCosmicRayAcceleration2019, condorelliImpactGalaxyClusters2023b}. These UHECRs rapidly lose energy through Bethe-Heitler pair production interactions with the Cosmic Microwave Background Radiation (CMBR), generating high-energy electron-positron pairs. Subsequently, these electrons quickly lose energy via the IC and synchrotron processes, leaving a $\gamma$-ray signature near the site of UHECR acceleration. In such cases, the emission would follow the spatial distribution of the shocks and present annulus/ring-type morphologies. However, we found no indications of such $\gamma$-ray emissions in the current LHAASO skymaps.

Based on this physical process, in addition to the previous constraints derived from the $pp$ interaction on CRp at tens of TeV, we can set limits on much higher-energy CRs, reaching up to $\sim 10^{18}-10^{19}$ eV. The $\gamma$-ray emission in this scenario reveals a quite hard spectrum, with a peak in the intrinsic spectrum at $10-100~\rm TeV$. This makes LHAASO an ideal instrument for such a study.

In this work, we compared the upper limit we derived with the calculations in \citet{vannoniAccelerationRadiationUltrahigh2011}, where both linear and nonlinear diffusive shock acceleration (DSA) were considered for a fixed total cosmic-ray energy. By comparing Figures 7 and 13 from this study with the upper limits we provided, we further constrained the UHECRs accelerated in galaxy clusters. The solid and dashed gray lines on the left-hand side of Figure \ref{fig:fig2} respectively show the expected $\gamma$-ray spectra from linear and nonlinear DSA under the aforementioned physical assumptions, scaled to match our most stringent constraints. It should be noted that the upper limit we derived in this work is the intrinsic $\gamma$-ray spectrum before absorption, and it was also compared with the intrinsic $\gamma$-ray spectra calculated in \citet{vannoniAccelerationRadiationUltrahigh2011}.

For linear DSA, we constrained $X_{CRp}$ to be less than 0.12, 0.08, and 0.12 for the Coma, Perseus, and Virgo clusters, respectively. For nonlinear DSA, these limits were tightened to 0.011, 0.008, and 0.011. Detailed information is also provided in Table \ref{tab:xcr}. The tighter constraints in the non-linear DSA scenario are expected, as shock modification produces a harder CR spectrum \citep{vannoniAccelerationRadiationUltrahigh2011}. As a result, the available kinetic energy transferred to relativistic particles, is accumulated at the highest energies. These high-energy particles dominate the contribution to the pair production and, ultimately, to the IC $\gamma$-ray emissions of the secondary electrons.



It is noteworthy that the propagation effects of UHECRs cannot be entirely disregarded. While \citet{vannoniAccelerationRadiationUltrahigh2011} did not consider the propagation and escape processes of UHECRs in galaxy clusters, recent simulation studies \citep{condorelliImpactGalaxyClusters2023b} show that approximately 40\% of $10^{19}\,\text{eV}$ cosmic rays escape from cluster environments with central magnetic fields of $3\,\mu\text{G}$. This significant escape fraction may substantially modify the predicted $\gamma$-ray spectrum. The non-detection of characteristic annulus/ring-type $\gamma$-ray features in LHAASO observations under the $p\gamma$+IC scenario may imply two possibilities: either the UHECR acceleration efficiency is insufficient to reach the required energy densities or UHECRs ($E \gtrsim 10^{18}\,\text{eV}$) have escaped from the cluster environment. We stress that our current conclusions are conditional on the crucial assumption of effective UHECR confinement within galaxy clusters. A comprehensive consideration of particle escape and its energy-dependent effects will be indispensable for future modeling efforts.


To summarize, by leveraging the upper limit of $\gamma$-rays in nearby galaxy clusters, we have derived stringent constraints on relativistic particles above 10 TeV in these systems. Additionally, we also establish stringent upper limit on the contents of UHECRs in these systems, through the $\gamma$-rays generated by secondary electrons via the $p\gamma$ process. The accumulation of LHAASO data in the next few decades will further deepen our understanding of CRs in galactic clusters and the origin of UHECRs.

In this field, multi-messenger observations, such as those involving neutrinos, can also provide important information. Moreover, the absorption of neutrinos is negligible, which gives them a distinct advantage at TeV energies and above. For example, \citet{shiConstrainingBaryonLoading2023a} utilized IceCube data to constrain the average baryon loading factor in galaxy cluster populations. However, due to the current sensitivity limitations of neutrino detectors, these constraints are not yet as stringent as those obtained from $\gamma$-ray observations. The forthcoming next-generation neutrino detectors, such as KM3NET\citep{margiottaKM3NeTDeepseaNeutrino2014}, ICECUBE Gen-2\citep{aartsenIceCubeGen2WindowExtreme2021}, Trident\citep{yeMulticubickilometreNeutrinoTelescope2023} and HUNT\citep{huangProposalHighEnergy2023}, will undoubtedly provide more informative observations and offer deeper insights into this field.



\acknowledgments

We would like to thank all staff members who work at the LHAASO site above 4400 meter above
the sea level year round to maintain the detector and keep the water recycling system, electricity power supply and other components of the experiment operating smoothly. We are grateful to Chengdu Management Committee of Tianfu New Area for the constant financial support for research with LHAASO data. We appreciate the computing and data service support provided by the National High Energy Physics Data Center for the data analysis in this paper. This research work is supported by the following grants: The National Natural Science Foundation of China No.12393854, No.12393851, No.12393852, No.12393853, No.12205314, No.12105301, No.12305120, No.12261160362, No.12105294, No.U1931201, No.12375107, No.12173039, the Department of Science and Technology of Sichuan Province, China No.24NSFSC2319, Project for Young Scientists in Basic Research of Chinese Academy of Sciences No.YSBR-061, and in Thailand by the National Science and Technology Development Agency (NSTDA) and the National Research Council of Thailand (NRCT) under the High-Potential Research Team Grant Program (N42A650868). 

\bibliography{cite}{}

@article{ferettiClustersGalaxiesObservational2012,
  title = {Clusters of Galaxies : Observational Properties of the Diffuse Radio Emission},
  shorttitle = {Clusters of Galaxies},
  author = {Feretti, Luigina and Giovannini, Gabriele and Govoni, Federica and Murgia, Matteo},
  year = {2012},
  month = oct,
  journal = {The Astronomy and Astrophysics Review},
  volume = {20},
  number = {1},
  eprint = {1205.1919},
  primaryclass = {astro-ph},
  pages = {54},
  issn = {0935-4956, 1432-0754},
  doi = {10.1007/s00159-012-0054-z},
  urldate = {2024-04-05},
  archiveprefix = {arxiv},
  lccn = {1}
}

@article{brunettiCosmicRaysGalaxy2014,
  title = {Cosmic Rays in Galaxy Clusters and Their Non-Thermal Emission},
  author = {Brunetti, G. and Jones, T. W.},
  year = {2014},
  month = apr,
  journal = {International Journal of Modern Physics D},
  volume = {23},
  number = {04},
  eprint = {1401.7519},
  primaryclass = {astro-ph},
  pages = {1430007},
  issn = {0218-2718, 1793-6594},
  doi = {10.1142/S0218271814300079},
  urldate = {2024-04-05},
  archiveprefix = {arxiv},
  lccn = {4}
}

@article{govoniMAGNETICFIELDSCLUSTERS2004,
  title = {{{MAGNETIC FIELDS IN CLUSTERS OF GALAXIES}}},
  author = {Govoni, Federica and Feretti, Luigina},
  year = {2004},
  doi = {10.1142/S0218271804005080},
  langid = {english}
}

@article{reimerEGRETUpperLimits2003,
  title = {{{EGRET}} Upper Limits on the High-Energy Gamma-Ray Emission of Galaxy Clusters},
  author = {Reimer, O. and Pohl, M. and Sreekumar, P. and Mattox, J. R.},
  year = {2003},
  month = may,
  journal = {The Astrophysical Journal},
  volume = {588},
  number = {1},
  eprint = {astro-ph/0301362},
  pages = {155--164},
  issn = {0004-637X, 1538-4357},
  doi = {10.1086/374046},
  urldate = {2024-04-06},
  archiveprefix = {arxiv},
  lccn = {2}
}

@article{perkinsTeVGammaRay2006,
  title = {{{TeV Gamma}}-{{Ray Observations}} of the {{Perseus}} and {{Abell}} 2029 {{Galaxy Clusters}}},
  author = {Perkins, J. S. and Badran, H. M. and Blaylock, G. and Bradbury, S. M. and Cogan, P. and Chow, Y. C. K. and Cui, W. and Daniel, M. K. and Falcone, A. D. and Fegan, S. J. and Finley, J. P. and Fortin, P. and Fortson, L. F. and Gillanders, G. H. and Gutierrez, K. J. and Grube, J. and Hall, J. and Hanna, D. and Holder, J. and Horan, D. and Hughes, S. B. and Humensky, T. B. and Kenny, G. E. and Kertzman, M. and Kieda, D. B. and Kildea, J. and Kosack, K. and Krawczynski, H. and Krennrich, F. and Lang, M. J. and LeBohec, S. and Maier, G. and Moriarty, P. and Ong, R. A. and Pohl, M. and Ragan, K. and Rebillot, P. F. and Sembroski, G. H. and Steele, D. and Swordy, S. P and Valcarcel, L. and Vassiliev, V. V. and Wakely, S. P. and Weekes, T. C. and Williams, D. A. and {(The VERITAS Collaboration)}},
  year = {2006},
  month = jun,
  journal = {The Astrophysical Journal},
  volume = {644},
  number = {1},
  pages = {148--154},
  issn = {0004-637X, 1538-4357},
  doi = {10.1086/503321},
  urldate = {2024-04-07},
  langid = {english},
  lccn = {2}
}

@article{xiDetectionGammarayEmission2018,
  title = {Detection of Gamma-Ray Emission from the {{Coma}} Cluster with {{Fermi Large Area Telescope}} and Tentative Evidence for an Extended Spatial Structure},
  author = {Xi, Shao-Qiang and Wang, Xiang-Yu and Liang, Yun-Feng and Peng, Fang-Kun and Yang, Rui-Zhi and Liu, Ruo-Yu},
  year = {2018},
  month = sep,
  journal = {Physical Review D},
  volume = {98},
  number = {6},
  eprint = {1709.08319},
  primaryclass = {astro-ph},
  pages = {063006},
  issn = {2470-0010, 2470-0029},
  doi = {10.1103/PhysRevD.98.063006},
  urldate = {2023-06-20},
  archiveprefix = {arxiv},
  langid = {english},
  lccn = {2}
}

@article{collaborationConstrainingCosmicrayPressure2023c,
  title = {Constraining the Cosmic-Ray Pressure in the Inner {{Virgo Cluster}} Using {{H}}.{{E}}.{{S}}.{{S}}. Observations of {{M}} 87},
  author = {{HESS Collaboration} and Aharonian, F. and Benkhali, F. Ait and Arcaro, C. and Aschersleben, J. and Backes, M. and Martins, V. Barbosa and Batzofin, R. and Becherini, Y. and Berge, D. and Bernl{\"o}hr, K. and Bi, B. and B{\"o}ttcher, M. and Boisson, C. and Bolmont, J. and Borowska, J. and Bradascio, F. and Breuhaus, M. and Brose, R. and Brun, F. and Bruno, B. and Bulik, T. and {Burger-Scheidlin}, C. and Bylund, T. and Caroff, S. and Casanova, S. and Cecil, R. and Celic, J. and Cerruti, M. and Chand, T. and Chandra, S. and Chen, A. and Chibueze, J. and Chibueze, O. and Cotter, G. and Mbarubucyeye, J. Damascene and {Djannati-Ata{\"i}}, A. and Egberts, K. and Ernenwein, J.-P. and {de Clairfontaine}, G. Fichet and Filipovic, M. and Fontaine, G. and F{\"u}{\ss}ling, M. and Funk, S. and Gabici, S. and Ghafourizadeh, S. and Giavitto, G. and Glawion, D. and Glicenstein, J. F. and Goswami, P. and Grolleron, G. and Grondin, M.-H. and Haerer, L. and Haupt, M. and Hermann, G. and Hinton, J. A. and Holch, T. L. and Horns, D. and Jamrozy, M. and Jankowsky, F. and Joshi, V. and {Jung-Richardt}, I. and Kasai, E. and Katarzy{\'n}ski, K. and Khatoon, R. and Kh{\'e}lifi, B. and Klu{\'z}niak, W. and Komin, Nu and Kosack, K. and Kostunin, D. and Lang, R. G. and Stum, S. Le and Leitl, F. and Lemi{\`e}re, A. and {Lemoine-Goumard}, M. and Lenain, J.-P. and Leuschner, F. and Lohse, T. and Luashvili, A. and Lypova, I. and Mackey, J. and Malyshev, D. and Malyshev, D. and Marandon, V. and Marchegiani, P. and Marcowith, A. and Marinos, P. and {Mart{\'i}-Devesa}, G. and Marx, R. and Meyer, M. and Mitchell, A. and Moderski, R. and Mohrmann, L. and Montanari, A. and Moulin, E. and Muller, J. and Nakashima, K. and {de Naurois}, M. and Niemiec, J. and Noel, A. Priyana and O'Brien, P. and Ohm, S. and {Olivera-Nieto}, L. and Wilhelmi, E. de Ona and Panny, S. and Panter, M. and Parsons, R. D. and Peron, G. and Pita, S. and Prokhorov, D. A. and Prokoph, H. and P{\"u}hlhofer, G. and Quirrenbach, A. and Reichherzer, P. and Reimer, A. and Reimer, O. and Renaud, M. and Rieger, F. and Rowell, G. and Rudak, B. and Velasco, E. Ruiz and Sahakian, V. and Salzmann, H. and Sanchez, D. A. and Santangelo, A. and Sasaki, M. and Sch{\"a}fer, J. and Sch{\"u}ssler, F. and Schwanke, U. and Shapopi, J. N. S. and Sol, H. and Specovius, A. and Spencer, S. and Stawarz, {\L} and Steenkamp, R. and Steinmassl, S. and Steppa, C. and Sushch, I. and Suzuki, H. and Takahashi, T. and Tanaka, T. and Taylor, A. M. and Terrier, R. and Tsirou, M. and Tsuji, N. and Uchiyama, Y. and {van Eldik}, C. and {van Soelen}, B. and Vecchi, M. and Veh, J. and Venter, C. and Vink, J. and Wach, T. and Wagner, S. J. and White, R. and Wierzcholska, A. and Wong, Yu Wun and Zacharias, M. and Zargaryan, D. and Zdziarski, A. A. and Zech, A. and Zouari, S. and {\.Z}ywucka, N.},
  year = {2023},
  month = jul,
  journal = {Astronomy \& Astrophysics},
  volume = {675},
  eprint = {2305.09607},
  primaryclass = {astro-ph},
  pages = {A138},
  issn = {0004-6361, 1432-0746},
  doi = {10.1051/0004-6361/202346056},
  urldate = {2023-09-11},
  archiveprefix = {arxiv},
  langid = {english},
  lccn = {2}
}

@article{thehesscollaborationConstraintsMultiTeVParticle2009,
  title = {Constraints on the Multi-{{TeV}} Particle Population in the {{Coma Galaxy Cluster}} with {{H}}.{{E}}.{{S}}.{{S}}. Observations},
  author = {{HESS Collaboration} and Aharonian, F. A.},
  year = {2009},
  month = aug,
  journal = {Astronomy \& Astrophysics},
  volume = {502},
  number = {2},
  eprint = {0907.0727},
  primaryclass = {astro-ph},
  pages = {437--443},
  issn = {0004-6361, 1432-0746},
  doi = {10.1051/0004-6361/200912086},
  urldate = {2023-04-14},
  archiveprefix = {arxiv},
  lccn = {2}
}

@article{arlenCONSTRAINTSCOSMICRAYS2012,
  title = {{{CONSTRAINTS ON COSMIC RAYS}}, {{MAGNETIC FIELDS}}, {{AND DARK MATTER FROM GAMMA-RAY OBSERVATIONS OF THE COMA CLUSTER OF GALAXIES WITH VERITAS AND}} {{{\emph{FERMI}}}}},
  author = {Arlen, T. and Aune, T. and Beilicke, M. and Benbow, W. and Bouvier, A. and Buckley, J. H. and Bugaev, V. and Byrum, K. and Cannon, A. and Cesarini, A. and Ciupik, L. and {Collins-Hughes}, E. and Connolly, M. P. and Cui, W. and Dickherber, R. and Dumm, J. and Falcone, A. and Federici, S. and Feng, Q. and Finley, J. P. and Finnegan, G. and Fortson, L. and Furniss, A. and Galante, N. and Gall, D. and Godambe, S. and Griffin, S. and Grube, J. and Gyuk, G. and Holder, J. and Huan, H. and Hughes, G. and Humensky, T. B. and Imran, A. and Kaaret, P. and Karlsson, N. and Kertzman, M. and Khassen, Y. and Kieda, D. and Krawczynski, H. and Krennrich, F. and Lee, K. and Madhavan, A. S and Maier, G. and Majumdar, P. and McArthur, S. and McCann, A. and Moriarty, P. and Mukherjee, R. and Nelson, T. and O'Faol{\'a}in De Bhr{\'o}it, A. and Ong, R. A. and Orr, M. and Otte, A. N. and Park, N. and Perkins, J. S. and Pohl, M. and Prokoph, H. and Quinn, J. and Ragan, K. and Reyes, L. C. and Reynolds, P. T. and Roache, E. and Ruppel, J. and Saxon, D. B. and Schroedter, M. and Sembroski, G. H. and Skole, C. and Smith, A. W. and Telezhinsky, I. and Te{\v s}i{\'c}, G. and Theiling, M. and Thibadeau, S. and Tsurusaki, K. and Varlotta, A. and Vivier, M. and Wakely, S. P. and Ward, J. E. and Weinstein, A. and Welsing, R. and Williams, D. A. and Zitzer, B. and Pfrommer, C. and Pinzke, A.},
  year = {2012},
  month = oct,
  journal = {The Astrophysical Journal},
  volume = {757},
  number = {2},
  pages = {123},
  issn = {0004-637X, 1538-4357},
  doi = {10.1088/0004-637X/757/2/123},
  urldate = {2023-06-20},
  langid = {english},
  lccn = {2}
}

@article{colafrancescoClustersGalaxiesDiffuse1998,
  title = {Clusters of Galaxies and the Diffuse Gamma-Ray Background},
  author = {Colafrancesco, Sergio and Blasi, Pasquale},
  year = {1998},
  month = oct,
  journal = {Astroparticle Physics},
  volume = {9},
  pages = {227--246},
  issn = {0927-6505},
  doi = {10.1016/S0927-6505(98)00018-8},
  urldate = {2024-04-07},
  lccn = {3}
}

@article{ryuCosmologicalShockWaves2003,
  title = {Cosmological {{Shock Waves}} and {{Their Role}} in the {{Large-Scale Structure}} of the {{Universe}}},
  author = {Ryu, Dongsu and Kang, Hyesung and Hallman, Eric and Jones, T. W.},
  year = {2003},
  month = aug,
  journal = {The Astrophysical Journal},
  volume = {593},
  pages = {599--610},
  issn = {0004-637X},
  doi = {10.1086/376723},
  urldate = {2024-04-07},
  lccn = {2}
}

@article{brunettiAlfvenicReaccelerationRelativistic2005,
  title = {Alfv{\'e}nic Reacceleration of Relativistic Particles in Galaxy Clusters in the Presence of Secondary Electrons and Positrons: {{Alfv{\'e}nic}} Reacceleration in Galaxy Clusters},
  shorttitle = {Alfv{\'e}nic Reacceleration of Relativistic Particles in Galaxy Clusters in the Presence of Secondary Electrons and Positrons},
  author = {Brunetti, G. and Blasi, P.},
  year = {2005},
  month = sep,
  journal = {Monthly Notices of the Royal Astronomical Society},
  volume = {363},
  number = {4},
  pages = {1173--1187},
  issn = {00358711},
  doi = {10.1111/j.1365-2966.2005.09511.x},
  urldate = {2024-04-07},
  langid = {english},
  lccn = {2}
}

@article{hintonRayEmissionAssociated2007,
  title = {-Ray Emission Associated with Cluster-Scale {{AGN}} Outbursts},
  author = {Hinton, J. A. and Domainko, W. and Pope, E. C. D.},
  year = {2007},
  month = nov,
  journal = {Monthly Notices of the Royal Astronomical Society},
  volume = {382},
  number = {1},
  pages = {466--472},
  issn = {0035-8711, 1365-2966},
  doi = {10.1111/j.1365-2966.2007.12395.x},
  urldate = {2024-04-07},
  langid = {english},
  lccn = {2}
}

@article{dennisonFormationRadioHalos1980,
  title = {Formation of Radio Halos in Clusters of Galaxies from Cosmic-Ray Protons.},
  author = {Dennison, B.},
  year = {1980},
  month = aug,
  journal = {The Astrophysical Journal},
  volume = {239},
  pages = {L93-L96},
  issn = {0004-637X},
  doi = {10.1086/183300},
  urldate = {2024-04-07},
  lccn = {2}
}

@article{atoyanImplicationsNonthermalOrigin2000,
  title = {Implications of a {{Nonthermal Origin}} of the {{Excess Extreme-Ultraviolet Emission}} from the {{Coma Cluster}} of {{Galaxies}}},
  author = {Atoyan, A. M. and V{\"o}lk, H. J.},
  year = {2000},
  month = may,
  journal = {The Astrophysical Journal},
  volume = {535},
  pages = {45--52},
  issn = {0004-637X},
  doi = {10.1086/308828},
  urldate = {2024-04-07},
  lccn = {2}
}

@article{kelnerEnergySpectraGamma2008,
  title = {Energy Spectra of Gamma Rays, Electrons, and Neutrinos Produced at Interactions of Relativistic Protons with Low Energy Radiation},
  author = {Kelner, S. R. and Aharonian, F. A.},
  year = {2008},
  month = aug,
  journal = {Physical Review D},
  volume = {78},
  pages = {034013},
  issn = {1550-79980556-2821},
  doi = {10.1103/PhysRevD.78.034013},
  urldate = {2024-04-08},
  lccn = {2}
}

@article{giovanniniHaloRadioSource1993,
  title = {The {{Halo Radio Source Coma C}} and the {{Origin}} of {{Halo Sources}}},
  author = {Giovannini, G. and Feretti, L. and Venturi, T. and Kim, K. -T. and Kronberg, P. P.},
  year = {1993},
  month = apr,
  journal = {The Astrophysical Journal},
  volume = {406},
  pages = {399},
  issn = {0004-637X},
  doi = {10.1086/172451},
  urldate = {2024-04-08},
  lccn = {2}
}

@article{planckcollaborationPlanckIntermediateResults2013,
  title = {Planck Intermediate Results. {{X}}. {{Physics}} of the Hot Gas in the {{Coma}} Cluster},
  author = {{Planck Collaboration} and Ade, P. A. R. and Aghanim, N. and Arnaud, M. and Ashdown, M. and {Atrio-Barandela}, F. and Aumont, J. and Baccigalupi, C. and Balbi, A. and Banday, A. J. and Barreiro, R. B. and Bartlett, J. G. and Battaner, E. and Benabed, K. and Beno{\^i}t, A. and Bernard, J. -P. and Bersanelli, M. and Bikmaev, I. and B{\"o}hringer, H. and Bonaldi, A. and Bond, J. R. and Borrill, J. and Bouchet, F. R. and Bourdin, H. and Brown, M. L. and Brown, S. D. and Burenin, R. and Burigana, C. and Cabella, P. and Cardoso, J. -F. and Carvalho, P. and Catalano, A. and Cay{\'o}n, L. and Chiang, L. -Y. and Chon, G. and Christensen, P. R. and Churazov, E. and Clements, D. L. and Colafrancesco, S. and Colombo, L. P. L. and Coulais, A. and Crill, B. P. and Cuttaia, F. and Da Silva, A. and Dahle, H. and Danese, L. and Davis, R. J. and {de Bernardis}, P. and {de Gasperis}, G. and {de Rosa}, A. and {de Zotti}, G. and Delabrouille, J. and D{\'e}mocl{\`e}s, J. and D{\'e}sert, F. -X. and Dickinson, C. and Diego, J. M. and Dolag, K. and Dole, H. and Donzelli, S. and Dor{\'e}, O. and D{\"o}rl, U. and Douspis, M. and Dupac, X. and En{\ss}lin, T. A. and Eriksen, H. K. and Finelli, F. and {Flores-Cacho}, I. and Forni, O. and Frailis, M. and Franceschi, E. and Frommert, M. and Galeotta, S. and Ganga, K. and {G{\'e}nova-Santos}, R. T. and Giard, M. and Gilfanov, M. and {Gonz{\'a}lez-Nuevo}, J. and G{\'o}rski, K. M. and Gregorio, A. and Gruppuso, A. and Hansen, F. K. and Harrison, D. and {Henrot-Versill{\'e}}, S. and {Hern{\'a}ndez-Monteagudo}, C. and Hildebrandt, S. R. and Hivon, E. and Hobson, M. and Holmes, W. A. and Hornstrup, A. and Hovest, W. and Huffenberger, K. M. and Hurier, G. and Jaffe, T. R. and Jagemann, T. and Jones, W. C. and Juvela, M. and Keih{\"a}nen, E. and Khamitov, I. and Kneissl, R. and Knoche, J. and Knox, L. and Kunz, M. and {Kurki-Suonio}, H. and Lagache, G. and L{\"a}hteenm{\"a}ki, A. and Lamarre, J. -M. and Lasenby, A. and Lawrence, C. R. and Le Jeune, M. and Leonardi, R. and Lilje, P. B. and {Linden-V{\o}rnle}, M. and {L{\'o}pez-Caniego}, M. and Lubin, P. M. and {Mac{\'i}as-P{\'e}rez}, J. F. and Maffei, B. and Maino, D. and Mandolesi, N. and Maris, M. and Marleau, F. and {Mart{\'i}nez-Gonz{\'a}lez}, E. and Masi, S. and Massardi, M. and Matarrese, S. and Matthai, F. and Mazzotta, P. and Mei, S. and Melchiorri, A. and Melin, J. -B. and Mendes, L. and Mennella, A. and Mitra, S. and {Miville-Desch{\^e}nes}, M. -A. and Moneti, A. and Montier, L. and Morgante, G. and Munshi, D. and Murphy, J. A. and Naselsky, P. and Natoli, P. and {N{\o}rgaard-Nielsen}, H. U. and Noviello, F. and Novikov, D. and Novikov, I. and Osborne, S. and Pajot, F. and Paoletti, D. and Perdereau, O. and Perrotta, F. and Piacentini, F. and Piat, M. and Pierpaoli, E. and Piffaretti, R. and Plaszczynski, S. and Pointecouteau, E. and Polenta, G. and Ponthieu, N. and Popa, L. and Poutanen, T. and Pratt, G. W. and Prunet, S. and Puget, J. -L. and Rachen, J. P. and Rebolo, R. and Reinecke, M. and Remazeilles, M. and Renault, C. and Ricciardi, S. and Riller, T. and Ristorcelli, I. and Rocha, G. and Roman, M. and Rosset, C. and Rossetti, M. and {Rubi{\~n}o-Mart{\'i}n}, J. A. and Rudnick, L. and Rusholme, B. and Sandri, M. and Savini, G. and Schaefer, B. M. and Scott, D. and Smoot, G. F. and Stivoli, F. and Sudiwala, R. and Sunyaev, R. and Sutton, D. and {Suur-Uski}, A. -S. and Sygnet, J. -F. and Tauber, J. A. and Terenzi, L. and Toffolatti, L. and Tomasi, M. and Tristram, M. and Tuovinen, J. and T{\"u}rler, M. and Umana, G. and Valenziano, L. and Van Tent, B. and Varis, J. and Vielva, P. and Villa, F. and Vittorio, N. and Wade, L. A. and Wandelt, B. D. and Welikala, N. and White, S. D. M. and Yvon, D. and Zacchei, A. and Zaroubi, S. and Zonca, A.},
  year = {2013},
  month = jun,
  journal = {Astronomy and Astrophysics},
  volume = {554},
  pages = {A140},
  issn = {0004-6361},
  doi = {10.1051/0004-6361/201220247},
  urldate = {2024-04-08}
}

@article{simionescuWitnessingGrowthNearest2017,
  title = {Witnessing the Growth of the Nearest Galaxy Cluster: Thermodynamics of the {{Virgo Cluster}} Outskirts},
  shorttitle = {Witnessing the Growth of the Nearest Galaxy Cluster},
  author = {Simionescu, A. and Werner, N. and Mantz, A. and Allen, S. W. and Urban, O.},
  year = {2017},
  month = aug,
  journal = {Monthly Notices of the Royal Astronomical Society},
  volume = {469},
  number = {2},
  pages = {1476--1495},
  issn = {0035-8711},
  doi = {10.1093/mnras/stx919},
  urldate = {2024-04-08},
  lccn = {2}
}

@article{thierbachDiffuseRadioEmission2003,
  title = {The Diffuse Radio Emission from the {{Coma}} Cluster at 2.675 {{GHz}} and 4.85 {{GHz}}},
  author = {Thierbach, M. and Klein, U. and Wielebinski, R.},
  year = {2003},
  month = jan,
  journal = {Astronomy and Astrophysics},
  volume = {397},
  pages = {53--61},
  issn = {0004-6361},
  doi = {10.1051/0004-6361:20021474},
  urldate = {2024-04-08}
}

@misc{consortiumProspectsGammaRay2023,
  title = {Prospects for \${\textbackslash}gamma\$-Ray Observations of the {{Perseus}} Galaxy Cluster with the {{Cherenkov Telescope Array}}},
  author = {{CTA Consortium} and Abe, K. and Abe, S. and Acero, F. and Acharyya, A. and Adam, R. and {Aguasca-Cabot}, A. and Agudo, I. and {Aguirre-Santaella}, A. and Alfaro, J. and Alfaro, R. and {Alvarez-Crespo}, N. and Batista, R. Alves and Amans, J.-P. and Amato, E. and Ang{\"u}ner, E. O. and Antonelli, L. A. and Aramo, C. and Araya, M. and Arcaro, C. and Arrabito, L. and Asano, K. and Ascas{\'i}bar, Y. and Aschersleben, J. and Ashkar, H. and Stuani, L. Augusto and Baack, D. and Backes, M. and Baktash, A. and Balazs, C. and Balbo, M. and Ballester, O. and Larriva, A. Baquero and Martins, V. Barbosa and {de Almeida}, U. Barres and Barrio, J. A. and Batista, P. I. and Batkovic, I. and Batzofin, R. and Baxter, J. and Gonz{\'a}lez, J. Becerra and Beck, G. and Tjus, J. Becker and Benbow, W. and Medrano, J. Bernete and Bernl{\"o}hr, K. and Berti, A. and Bertucci, B. and Beshley, V. and Bhattacharjee, P. and Bhattacharyya, S. and Bi, B. and Biederbeck, N. and Biland, A. and Bissaldi, E. and Biteau, J. and Blanch, O. and Blazek, J. and Boisson, C. and Bolmont, J. and Bordas, P. and Bosnjak, Z. and Bottacini, E. and Bradascio, F. and Braiding, C. and Bronzini, E. and Brose, R. and Brown, A. M. and Brun, F. and Brunetti, G. and Bucciantini, N. and Bulgarelli, A. and Burelli, I. and Burmistrov, L. and Burton, M. and Bylund, T. and Calisse, P. G. and {Campoy-Ordaz}, A. and Cantlay, B. K. and Capalbi, M. and Caproni, A. and {Capuzzo-Dolcetta}, R. and Caraveo, P. and Caroff, S. and Carosi, R. and Carquin, E. and Carrasco, M.-S. and Cascone, E. and Cassol, F. and {Castro-Tirado}, A. J. and Cerasole, D. and Cerruti, M. and Chadwick, P. and Chaty, S. and Chen, A. W. and Chernyakova, M. and Chiavassa, A. and Chudoba, J. and Chytka, L. and Cifuentes, A. and Araujo, C. H. Coimbra and Conforti, V. and Conte, F. and Contreras, J. L. and Cortina, J. and Costa, A. and Costantini, H. and Cotter, G. and Cristofari, P. and Cuevas, O. and {Curtis-Ginsberg}, Z. and D'Amico, G. and D'Ammando, F. and Dalchenko, M. and Dazzi, F. and {de Lavergne}, M. de Bony and De Caprio, V. and Laadim, F. De Frondat and Pino, E. M. de Gouveia Dal and De Lotto, B. and De Lucia, M. and De Martino, D. and {de Menezes}, R. and {de Naurois}, M. and De Simone, N. and {de Souza}, V. and {del Valle}, M. V. and Delagnes, E. and Giler, A. G. Delgado and Delgado, C. and Dell'aiera, M. and {della Volpe}, D. and Depaoli, D. and Di Girolamo, T. and Di Piano, A. and Di Pierro, F. and Di Tria, R. and Di Venere, L. and Diebold, S. and {Djannati-Ata{\"i}}, A. and Djuvsland, J. and Dominik, R. M. and Donini, A. and Dorner, D. and D{\"o}rner, J. and Doro, M. and dos Anjos, R. D. C. and Dournaux, J.-L. and Duangchan, C. and Dubos, C. and Dumora, D. and Dwarkadas, V. V. and Ebr, J. and Eckner, C. and Egberts, K. and Einecke, S. and Els{\"a}sser, D. and Emery, G. and Godoy, M. Escobar and Escudero, J. and Esposito, P. and Ettori, S. and Evoli, C. and {Falceta-Goncalves}, D. and Ramazani, V. Fallah and Fattorini, A. and Faure, A. and Fedorova, E. and Fegan, S. and Feijen, K. and Feng, Q. and Ferrand, G. and Ferrarotto, F. and Fiandrini, E. and Fiasson, A. and Filipovic, M. and Fioretti, V. and Foffano, L. and Guiteras, L. Font and Fontaine, G. and Fr{\"o}se, S. and Fukazawa, Y. and Fukui, Y. and Gaggero, D. and Galanti, G. and Gallozzi, S. and Gammaldi, V. and Garczarczyk, M. and Gasbarra, C. and Gasparrini, D. and Gaug, M. and Ghalumyan, A. and Gianotti, F. and Giarrusso, M. and Giesbrecht, J. and Giglietto, N. and Giordano, F. and Glicenstein, J.-F. and G{\"o}ksu, H. and Goldoni, P. and Gonz{\'a}lez, J. M. and Gonz{\'a}lez, M. M. and Coelho, J. Goulart and Granot, J. and Grau, R. and Gr{\'e}aux, L. and Green, D. and Green, J. G. and Grenier, I. and Grolleron, G. and Grube, J. and Gueta, O. and Hackfeld, J. and Hadasch, D. and Hamal, P. and Hanlon, W. and Hara, S. and Harvey, V. M. and Hassan, T. and Heckmann, L. and Heller, M. and Cadena, S. Hern{\'a}ndez and Hervet, O. and Hie, J. and Hiroshima, N. and Hnatyk, B. and Hnatyk, R. and Hoang, J. and Hoffmann, D. and Hofmann, W. and Holder, J. and Horan, D. and Horvath, P. and Hrupec122, D. and H{\"u}tten, M. and Iarlori, M. and Inada, T. and Incardona, F. and Inoue, S. and Iocco, F. and Iori, M. and Jamrozy, M. and Janecek, P. and Jankowsky, F. and Jarnot, C. and Jean, P. and Mart{\'i}nez, I. Jim{\'e}nez and Jin, W. and {Juramy-Gilles}, C. and Jurysek, J. and Kagaya, M. and Kantzas, D. and Karas, V. and Katagiri, H. and Kataoka, J. and Kaufmann, S. and Kerszberg, D. and Kh{\'e}lifi, B. and Kissmann, R. and Kleiner, T. and Kluge, G. and Klu{\'z}niak, W. and Kn{\"o}dlseder, J. and Kobayashi, Y. and Kohri, K. and Komin, N. and Kornecki, P. and Kosack, K. and Kowal, G. and Kubo, H. and Kushida, J. and La Barbera, A. and La Palombara, N. and L{\'a}inez, M. and Lamastra, A. and Lapington, J. and Laporte, P. and Lazarevi{\'c}, S. and Leitgeb, F. and {Lemoine-Goumard}, M. and Lenain, J.-P. and Leone, F. and Leto, G. and Leuschner, F. and Lindfors, E. and Linhoff, M. and Liodakis, I. and Lombardi, S. and Longo, F. and {L{\'o}pez-Coto}, R. and {L{\'o}pez-Moya}, M. and {L{\'o}pez-Oramas}, A. and Loporchio, S. and {Luque-Escamilla}, P. L. and Macias, O. and Mackey, J. and Majumdar, P. and Malyshev, D. and Mandat, D. and Manganaro, M. and Manic{\`o}, G. and Mariotti, M. and Markoff, S. and M{\'a}rquez, I. and Marquez, P. and Marsella, G. and Mart{\'i}nez, G. A. and Mart{\'i}nez, M. and Martinez, O. and Marty, C. and {Mas-Aguilar}, A. and Mastropietro, M. and Maurin, G. and Mazin, D. and Melkumyan, D. and Mello, A. J. T. S. and Meunier, J.-L. and Meyer, D. M.-A. and Meyer, M. and Miceli, D. and Michailidis, M. and Micha{\l}owski, J. and Miener, T. and Miranda, J. M. and Mitchell, A. and Mizote, M. and Mizuno, T. and Moderski, R. and Molero, M. and Molfese, C. and Molina, E. and Montaruli, T. and Morcuende, D. and Morik, K. and Morlino, G. and Morselli, A. and Moulin, E. and Zamanillo, V. Moya and Munari, K. and Murach, T. and Muraczewski, A. and Muraishi, H. and Nagataki, S. and Nakamori, T. and Nemmen, R. and Neyroud, N. and Nickel, L. and Niemiec, J. and Nieto, D. and Rosillo, M. Nievas and Niko{\l}ajuk, M. and Nishijima, K. and Noda, K. and Nosek, D. and Novotny, V. and Nozaki, S. and O'Brien, P. and Ohishi, M. and Ohtani, Y. and Okumura, A. and Olive, J.-F. and Olmi, B. and Ong, R. A. and Orienti, M. and Orito, R. and Orlandini, M. and Orlando, E. and Ostrowski, M. and Oya, I. and Pagliaro, A. and Palatiello, M. and Panebianco, G. and Paneque, D. and Pantaleo, F. R. and Paoletti, R. and Paredes, J. M. and Parmiggiani, N. and Patel, S. R. and Patricelli, B. and Pavlovi{\'c}, D. and Pech, M. and Pecimotika, M. and Pensec, U. and Peresano, M. and {P{\'e}rez-Romero}, J. and Peron, G. and Persic, M. and Petrucci, P.-O. and Petruk, O. and Piano, G. and Pierre, E. and Pietropaolo, E. and Pintore, F. and Pirola, G. and Pita, S. and Plard, C. and Podobnik, F. and Pohl, M. and Polo, M. and Pons, E. and Ponti, G. and Prandini, E. and Prast, J. and Principe, G. and Priyadarshi, C. and Produit, N. and Pueschel, E. and P{\"u}hlhofer, G. and Pumo, M. L. and Punch, M. and Queiroz, F. and Quirrenbach, A. and Rain{\`o}, S. and Rando, R. and Razzaque, S. and Recchia, S. and Regeard, M. and Reichherzer, P. and Reimer, A. and Reimer, O. and Reisenegger, A. and Rhode, W. and Ribeiro, D. and Rib{\'o}, M. and Richtler, T. and Rico, J. and Rieger, F. and Righi, C. and Riitano, L. and Rizi, V. and Roache, E. and Fernandez, G. Rodriguez and {Rodr{\'i}guez-V{\'a}zquez}, J. J. and Romano, P. and Romeo, G. and Rosado, J. and {de Leon}, A. Rosales and Rowell, G. and Rudak, B. and Rulten, C. B. and Russo, F. and Sadeh, I. and Saha, L. and Saito, T. and Salzmann, H. and Sanchez, D. and {S{\'a}nchez-Conde}, M. and Sangiorgi, P. and Sano, H. and Santander, M. and Santangelo, A. and {Santos-Lima}, R. and Sanuy, A. and {\v S}ari{\'c}, T. and Sarkar, A. and Sarkar, S. and Satalecka, K. and Saturni, F. G. and Savchenko, V. and Scherer, A. and Schipani, P. and Schleicher, B. and Schubert, J. L. and Schussler, F. and Schwanke, U. and Schwefer, G. and Arroyo, M. Seglar and Seiji, S. and Semikoz, D. and Sergijenko, O. and Servillat, M. and Sguera, V. and Shang, R. Y. and Sharma, P. and Siejkowski, H. and Sinha, A. and Siqueira, C. and Sliusar, V. and Slowikowska, A. and Sol, H. and Specovius, A. and Spencer, S. T. and Spiga, D. and Stamerra, A. and Stani{\v c}, S. and Starecki, T. and Starling, R. and Stawarz, {\L} and Steppa, C. and Stolarczyk, T. and Stri{\v s}kovi{\'c}, J. and Suda, Y. and Suomij{\"a}rvi, T. and Tajima, H. and Tak, D. and Takahashi, M. and Takeishi, R. and Tanaka, S. J. and Tavernier, T. and Tejedor, L. A. and Terauchi, K. and Terrier, R. and Teshima, M. and Tian, W. W. and Tibaldo, L. and Tibolla, O. and Torradeflot, F. and Torres, D. F. and Torresi, E. and Tosti, G. and Tosti, L. and Tothill, N. and Toussenel, F. and Touzard, V. and Tramacere, A. and Travnicek, P. and Tripodo, G. and Truzzi, S. and Tsiahina, A. and Tutone, A. and Vacula, M. and Vallage, B. and Vallania, P. and {van Eldik}, C. and {van Scherpenberg}, J. and Vandenbroucke, J. and Vassiliev, V. and Acosta, M. V{\'a}zquez and Vecchi, M. and Ventura, S. and Vercellone, S. and Verna, G. and Viana, A. and Viaux, N. and Vigliano, A. and Vigorito, C. F. and Vitale, V. and Vodeb, V. and Voisin, V. and Vorobiov, S. and Voutsinas, G. and Vovk, I. and Vuillaume, T. and Wagner, S. J. and Walter, R. and Wechakama, M. and White, R. and Wierzcholska, A. and Will, M. and Williams, D. A. and Wohlleben, F. and Wolter, A. and Yamamoto, T. and Yamazaki, R. and Yoshida, T. and Yoshikoshi, T. and Zacharias, M. and Zaharijas, G. and Zavrtanik, D. and Zavrtanik, M. and Zdziarski, A. A. and Zech, A. and Zhdanov, V. I. and {\v Z}ivec, M. and {Zuriaga-Puig}, J. and Luque, P. De la Torre},
  year = {2023},
  month = sep,
  number = {arXiv:2309.03712},
  eprint = {2309.03712},
  primaryclass = {astro-ph, physics:hep-ph},
  publisher = {arXiv},
  urldate = {2023-09-09},
  archiveprefix = {arxiv}
}

@article{vannoniAccelerationRadiationUltrahigh2011,
  title = {Acceleration and Radiation of Ultra-High Energy Protons in Galaxy Clusters},
  author = {Vannoni, G. and Aharonian, F. A. and Gabici, S. and Kelner, S. R. and Prosekin, A.},
  year = {2011},
  month = dec,
  journal = {Astronomy \& Astrophysics},
  volume = {536},
  eprint = {0910.5715},
  primaryclass = {astro-ph},
  pages = {A56},
  issn = {0004-6361, 1432-0746},
  doi = {10.1051/0004-6361/200913568},
  urldate = {2024-03-29},
  archiveprefix = {arxiv},
  lccn = {2}
}

@article{gendron-marsolaisHighresolutionVLALow2020,
  title = {High-Resolution {{VLA}} Low Radio Frequency Observations of the {{Perseus}} Cluster: Radio Lobes, Mini-Halo and Bent-Jet Radio Galaxies},
  shorttitle = {High-Resolution {{VLA}} Low Radio Frequency Observations of the {{Perseus}} Cluster},
  author = {{Gendron-Marsolais}, Marie-Lou and {Hlavacek-Larrondo}, Julie and {van Weeren}, Reinout J. and Rudnick, Lawrence and Clarke, Tracy E. and Sebastian, Biny and Mroczkowski, Tony and Fabian, Andrew C. and Blundell, Katherine M. and Sheldahl, Evan and Nyland, Kristina and Sanders, Jeremy S. and Peters, Wendy M. and Intema, Huib T.},
  year = {2020},
  month = nov,
  journal = {Monthly Notices of the Royal Astronomical Society},
  volume = {499},
  number = {4},
  eprint = {2005.12298},
  primaryclass = {astro-ph},
  pages = {5791--5805},
  issn = {0035-8711, 1365-2966},
  doi = {10.1093/mnras/staa2003},
  urldate = {2024-04-09},
  archiveprefix = {arxiv},
  lccn = {2}
}

@article{gendron-marsolaisVLAResolvesUnexpected2021,
  title = {{{VLA}} Resolves Unexpected Radio Structures in the {{Perseus}} Cluster of Galaxies},
  author = {{Gendron-Marsolais}, Marie-Lou and Hull, Charles L. H. and Perley, Rick and Rudnick, Lawrence and Kraft, Ralph and {Hlavacek-Larrondo}, Julie and Fabian, Andrew C. and Roediger, Elke and {van Weeren}, Reinout J. and {Richard-Laferri{\`e}re}, Annabelle and {Golden-Marx}, Emmet and Arakawa, Naoki and McBride, James D.},
  year = {2021},
  month = apr,
  journal = {The Astrophysical Journal},
  volume = {911},
  number = {1},
  eprint = {2101.05305},
  primaryclass = {astro-ph},
  pages = {56},
  issn = {0004-637X, 1538-4357},
  doi = {10.3847/1538-4357/abddbb},
  urldate = {2024-04-09},
  archiveprefix = {arxiv},
  lccn = {2}
}

@article{petersonXraySpectroscopyCooling2006,
  title = {X-Ray Spectroscopy of Cooling Clusters},
  author = {Peterson, J.R. and Fabian, A.C.},
  year = {2006},
  month = apr,
  journal = {Physics Reports},
  volume = {427},
  number = {1},
  pages = {1--39},
  issn = {03701573},
  doi = {10.1016/j.physrep.2005.12.007},
  urldate = {2024-04-09},
  copyright = {https://www.elsevier.com/tdm/userlicense/1.0/},
  langid = {english}
}

@article{cassanoConnectionGiantRadio2010,
  title = {On the Connection between Giant Radio Halos and Cluster Mergers},
  author = {Cassano, R. and Ettori, S. and Giacintucci, S. and Brunetti, G. and Markevitch, M. and Venturi, T. and Gitti, M.},
  year = {2010},
  month = oct,
  journal = {The Astrophysical Journal},
  volume = {721},
  number = {2},
  eprint = {1008.3624},
  primaryclass = {astro-ph},
  pages = {L82-L85},
  issn = {2041-8205, 2041-8213},
  doi = {10.1088/2041-8205/721/2/L82},
  urldate = {2024-04-09},
  archiveprefix = {arxiv},
  lccn = {2}
}

@article{fabianChandraImagingComplex2000,
  title = {Chandra Imaging of the Complex {{X-ray}} Core of the {{Perseus}} Cluster},
  author = {Fabian, A. C. and Sanders, J. S. and Ettori, S. and Taylor, G. B. and Allen, S. W. and Crawford, C. S. and Iwasawa, K. and Johnstone, R. M. and Ogle, P. M.},
  year = {2000},
  month = nov,
  journal = {Monthly Notices of the Royal Astronomical Society},
  volume = {318},
  number = {4},
  eprint = {astro-ph/0007456},
  pages = {L65-L68},
  issn = {0035-8711, 1365-2966},
  doi = {10.1046/j.1365-8711.2000.03904.x},
  urldate = {2024-04-09},
  archiveprefix = {arxiv},
  lccn = {2}
}

@article{churazovEvolutionBuoyantBubbles2001,
  title = {Evolution of {{Buoyant Bubbles}} in {{M87}}},
  author = {Churazov, E. and Br{\"u}ggen, M. and Kaiser, C. R. and B{\"o}hringer, H. and Forman, W.},
  year = {2001},
  month = jun,
  journal = {The Astrophysical Journal},
  volume = {554},
  number = {1},
  pages = {261},
  publisher = {IOP Publishing},
  issn = {0004-637X},
  doi = {10.1086/321357},
  urldate = {2024-04-09},
  langid = {english},
  lccn = {2}
}

@article{churazovTempestuousLifeR5002021,
  title = {Tempestuous Life beyond {{R500}}: {{X-ray}} View on the {{Coma}} Cluster with {{SRG}}/{{eROSITA}}. {{I}}. {{X-ray}} Morphology, Recent Merger, and Radio Halo Connection},
  shorttitle = {Tempestuous Life beyond {{R500}}},
  author = {Churazov, E. and Khabibullin, I. and Lyskova, N. and Sunyaev, R. and Bykov, A. M.},
  year = {2021},
  month = jul,
  journal = {Astronomy and Astrophysics},
  volume = {651},
  pages = {A41},
  issn = {0004-6361},
  doi = {10.1051/0004-6361/202040197},
  urldate = {2024-04-09}
}

@article{malavasiSpiderItsWeb2020,
  title = {Like a Spider in Its Web: A Study of the Large-Scale Structure around the {{Coma}} Cluster},
  shorttitle = {Like a Spider in Its Web},
  author = {Malavasi, Nicola and Aghanim, Nabila and Tanimura, Hideki and Bonjean, Victor and Douspis, Marian},
  year = {2020},
  month = feb,
  journal = {Astronomy \& Astrophysics},
  volume = {634},
  pages = {A30},
  publisher = {EDP Sciences},
  issn = {0004-6361, 1432-0746},
  doi = {10.1051/0004-6361/201936629},
  urldate = {2024-04-09},
  copyright = {{\copyright} N. Malavasi et al. 2020},
  langid = {english},
  lccn = {2}
}

@article{boltonVariableSourceRadio1948,
  title = {Variable {{Source}} of {{Radio Frequency Radiation}} in the {{Constellation}} of {{Cygnus}}},
  author = {Bolton, J. G. and Stanley, G. J.},
  year = {1948},
  month = feb,
  journal = {Nature},
  volume = {161},
  pages = {312--313},
  issn = {0028-0836},
  doi = {10.1038/161312b0},
  urldate = {2024-04-09},
  lccn = {1}
}

@article{vollmerDetectionRadioHalo2004,
  title = {Detection of a Radio Halo in the {{Virgo}} Cluster},
  author = {Vollmer, B. and Reich, W. and Wielebinski, R.},
  year = {2004},
  month = aug,
  journal = {Astronomy \& Astrophysics},
  volume = {423},
  number = {1},
  pages = {57--64},
  publisher = {EDP Sciences},
  issn = {0004-6361, 1432-0746},
  doi = {10.1051/0004-6361:20035783},
  urldate = {2024-04-08},
  copyright = {{\copyright} ESO, 2004},
  langid = {english},
  lccn = {2}
}

@article{baghmanyanDetailedStudyExtended2022,
  title = {Detailed Study of Extended Gamma-Ray Morphology in the Vicinity of the {{Coma}} Cluster with {{Fermi-LAT}}},
  author = {Baghmanyan, Vardan and Zargaryan, Davit and Aharonian, Felix and Yang, Ruizhi and Casanova, Sabrina and Mackey, Jonathan},
  year = {2022},
  month = aug,
  journal = {Monthly Notices of the Royal Astronomical Society},
  volume = {516},
  number = {1},
  eprint = {2110.00309},
  primaryclass = {astro-ph},
  pages = {562--571},
  issn = {0035-8711, 1365-2966},
  doi = {10.1093/mnras/stac2266},
  urldate = {2023-06-20},
  archiveprefix = {arxiv},
  langid = {english},
  lccn = {2}
}

@article{aleksicDetectionVeryhighEnergy2012,
  title = {Detection of Very-High Energy {$\gamma$}-Ray Emission from {{NGC}} 1275 by the {{MAGIC}} Telescopes},
  author = {Aleksi{\'c}, J. and Alvarez, E. A. and Antonelli, L. A. and Antoranz, P. and Asensio, M. and Backes, M. and de Almeida, U. Barres and Barrio, J. A. and Bastieri, D. and Gonz{\'a}lez, J. Becerra and Bednarek, W. and Berger, K. and Bernardini, E. and Biland, A. and Blanch, O. and Bock, R. K. and Boller, A. and Bonnoli, G. and Tridon, D. Borla and Bretz, T. and Ca{\~n}ellas, A. and Carmona, E. and Carosi, A. and Colin, P. and Colombo, E. and Contreras, J. L. and Cortina, J. and Cossio, L. and Covino, S. and Vela, P. Da and Dazzi, F. and Angelis, A. De and Caneva, G. De and del Pozo, E. De Cea and Lotto, B. De and Mendez, C. Delgado and Ortega, A. Diago and Doert, M. and Dom{\'i}nguez, A. and Prester, D. Dominis and Dorner, D. and Doro, M. and Eisenacher, D. and Elsaesser, D. and Ferenc, D. and Fonseca, M. V. and Font, L. and Fruck, C. and L{\'o}pez, R. J. Garc{\'i}a and Garczarczyk, M. and Garrido, D. and Giavitto, G. and Godinovi{\'c}, N. and Gozzini, S. R. and Hadasch, D. and H{\"a}fner, D. and Herrero, A. and Hildebrand, D. and {H{\"o}hne-M{\"o}nch}, D. and Hose, J. and Hrupec, D. and Huber, B. and Jogler, T. and Kadenius, V. and Kellermann, H. and Klepser, S. and Kr{\"a}henb{\"u}hl, T. and Krause, J. and Barbera, A. La and Lelas, D. and Leonardo, E. and Lewandowska, N. and Lindfors, E. and Lombardi, S. and L{\'o}pez, M. and {L{\'o}pez-Coto}, R. and {L{\'o}pez-Oramas}, A. and Lorenz, E. and Makariev, M. and Maneva, G. and Mankuzhiyil, N. and Mannheim, K. and Maraschi, L. and Mariotti, M. and Mart{\'i}nez, M. and Mazin, D. and Meucci, M. and Miranda, J. M. and Mirzoyan, R. and Mold{\'o}n, J. and Moralejo, A. and {Munar-Adrover}, P. and Niedzwiecki, A. and Nieto, D. and Nilsson, K. and Nowak, N. and Orito, R. and Paiano, S. and Paneque, D. and Paoletti, R. and Pardo, S. and Paredes, J. M. and Partini, S. and {Perez-Torres}, M. A. and Persic, M. and Peruzzo, L. and Pilia, M. and Pochon, J. and Prada, F. and Moroni, P. G. Prada and Prandini, E. and Gimenez, I. Puerto and Puljak, I. and Reichardt, I. and Reinthal, R. and Rhode, W. and Rib{\'o}, M. and Rico, J. and R{\"u}gamer, S. and Saggion, A. and Saito, K. and Saito, T. Y. and Salvati, M. and Satalecka, K. and Scalzotto, V. and Scapin, V. and Schultz, C. and Schweizer, T. and Shayduk, M. and Shore, S. N. and Sillanp{\"a}{\"a}, A. and Sitarek, J. and Snidaric, I. and Sobczynska, D. and Spanier, F. and Spiro, S. and Stamatescu, V. and Stamerra, A. and Steinke, B. and Storz, J. and Strah, N. and Sun, S. and Suri{\'c}, T. and Takalo, L. and Takami, H. and Tavecchio, F. and Temnikov, P. and Terzi{\'c}, T. and Tescaro, D. and Teshima, M. and Tibolla, O. and Torres, D. F. and Treves, A. and Uellenbeck, M. and Vogler, P. and Wagner, R. M. and Weitzel, Q. and Zabalza, V. and Zandanel, F. and Zanin, R. and Pfrommer, C. and Pinzke, A.},
  year = {2012},
  month = mar,
  journal = {Astronomy \& Astrophysics},
  volume = {539},
  pages = {L2},
  publisher = {EDP Sciences},
  issn = {0004-6361, 1432-0746},
  doi = {10.1051/0004-6361/201118668},
  urldate = {2024-04-14},
  copyright = {{\copyright} ESO, 2012},
  langid = {english},
  lccn = {2}
}

@article{neronovVeryHighenergyRay2010,
  title = {Very High-Energy {\emph{{$\gamma$}}} -Ray Emission from {{IC}} 310},
  author = {Neronov, A. and Semikoz, D. and Vovk, {\relax Ie}.},
  year = {2010},
  month = sep,
  journal = {Astronomy and Astrophysics},
  volume = {519},
  pages = {L6},
  issn = {0004-6361, 1432-0746},
  doi = {10.1051/0004-6361/201014499},
  urldate = {2024-04-14},
  langid = {english}
}

@article{aharonianGiantRadioGalaxy2003a,
  title = {Is the Giant Radio Galaxy {{M}} 87 a {{TeV}} Gamma-Ray Emitter?},
  author = {Aharonian, F. and Akhperjanian, A. and Beilicke, M. and Bernl{\"o}hr, K. and B{\"o}rst, H.-G. and Bojahr, H. and Bolz, O. and Coarasa, T. and Contreras, J. L. and Cortina, J. and Denninghoff, S. and Fonseca, M. V. and Girma, M. and G{\"o}tting, N. and Heinzelmann, G. and Hermann, G. and Heusler, A. and Hofmann, W. and Horns, D. and Jung, I. and Kankanyan, R. and Kestel, M. and Kohnle, A. and Konopelko, A. and Kornmeyer, H. and Kranich, D. and Lampeitl, H. and Lopez, M. and Lorenz, E. and Lucarelli, F. and Mang, O. and Meyer, H. and Mirzoyan, R. and Moralejo, A. and {Ona-Wilhelmi}, E. and Panter, M. and Plyasheshnikov, A. and P{\"u}hlhofer, G. and {R. De Los Reyes} and Rhode, W. and Ripken, J. and Rowell, G. and Sahakian, V. and Samorski, M. and Schilling, M. and Siems, M. and Sobzynska, D. and Stamm, W. and Tluczykont, M. and Vitale, V. and V{\"o}lk, H. J. and Wiedner, C. A. and Wittek, W.},
  year = {2003},
  month = may,
  journal = {Astronomy \& Astrophysics},
  volume = {403},
  number = {1},
  pages = {L1-L5},
  issn = {0004-6361, 1432-0746},
  doi = {10.1051/0004-6361:20030372},
  urldate = {2024-04-14},
  langid = {english},
  lccn = {2}
}

@article{adamRayDetectionComa2021,
  title = {{\emph{{$\gamma$}}} -Ray Detection toward the {{Coma}} Cluster with {{{\emph{Fermi}}}} -{{LAT}}: {{Implications}} for the Cosmic Ray Content in the Hadronic Scenario},
  shorttitle = {{\emph{{$\gamma$}}} -Ray Detection toward the {{Coma}} Cluster with {{{\emph{Fermi}}}} -{{LAT}}},
  author = {Adam, R. and Goksu, H. and Brown, S. and Rudnick, L. and Ferrari, C.},
  year = {2021},
  month = apr,
  journal = {Astronomy \& Astrophysics},
  volume = {648},
  pages = {A60},
  issn = {0004-6361, 1432-0746},
  doi = {10.1051/0004-6361/202039660},
  urldate = {2023-09-10},
  langid = {english},
  lccn = {2}
}

@article{magiccollaborationConstrainingCosmicRays2012,
  title = {Constraining {{Cosmic Rays}} and {{Magnetic Fields}} in the {{Perseus Galaxy Cluster}} with {{TeV}} Observations by the {{MAGIC}} Telescopes},
  author = {{MAGIC Collaboration} and Aleksi{\'c}, J. and Alvarez, E. A. and Antonelli, L. A. and Antoranz, P. and Asensio, M. and Backes, M. and {de Almeida}, U. Barres and Barrio, J. A. and Bastieri, D. and Gonz{\'a}lez, J. Becerra and Bednarek, W. and Berdyugin, A. and Berger, K. and Bernardini, E. and Biland, A. and Blanch, O. and Bock, R. K. and Boller, A. and Bonnoli, G. and Tridon, D. Borla and Braun, I. and Bretz, T. and Ca{\~n}ellas, A. and Carmona, E. and Carosi, A. and Colin, P. and Colombo, E. and Contreras, J. L. and Cortina, J. and Cossio, L. and Covino, S. and Dazzi, F. and De Angelis, A. and De Caneva, G. and {del Pozo}, E. De Cea and De Lotto, B. and Mendez, C. Delgado and Ortega, A. Diago and Doert, M. and Dom{\'i}nguez, A. and Prester, D. Dominis and Dorner, D. and Doro, M. and Eisenacher, D. and Elsaesser, D. and Ferenc, D. and Fonseca, M. V. and Font, L. and Fruck, C. and L{\'o}pez, R. J. Garc{\'i}a and Garczarczyk, M. and Garrido, D. and Giavitto, G. and Godinovi{\'c}, N. and Gozzini, S. R. and Hadasch, D. and H{\"a}fner, D. and Herrero, A. and Hildebrand, D. and {H{\"o}hne-M{\"o}nch}, D. and Hose, J. and Hrupec, D. and Jogler, T. and Kellermann, H. and Klepser, S. and Kr{\"a}henb{\"u}hl, T. and Krause, J. and Kushida, J. and La Barbera, A. and Lelas, D. and Leonardo, E. and Lewandowska, N. and Lindfors, E. and Lombardi, S. and L{\'o}pez, M. and {L{\'o}pez-Coto}, R. and {L{\'o}pez-Oramas}, A. and Lorenz, E. and Makariev, M. and Maneva, G. and Mankuzhiyil, N. and Mannheim, K. and Maraschi, L. and Mariotti, M. and Mart{\'i}nez, M. and Mazin, D. and Meucci, M. and Miranda, J. M. and Mirzoyan, R. and Mold{\'o}n, J. and Moralejo, A. and {Munar-Adrover}, P. and Niedzwiecki, A. and Nieto, D. and Nilsson, K. and Nowak, N. and Orito, R. and Paiano, S. and Paneque, D. and Paoletti, R. and Pardo, S. and Paredes, J. M. and Partini, S. and {Perez-Torres}, M. A. and Persic, M. and Peruzzo, L. and Pilia, M. and Pochon, J. and Prada, F. and Moroni, P. G. Prada and Prandini, E. and Gimenez, I. Puerto and Puljak, I. and Reichardt, I. and Reinthal, R. and Rhode, W. and Rib{\'o}, M. and Rico, J. and R{\"u}gamer, S. and Saggion, A. and Saito, K. and Saito, T. Y. and Salvati, M. and Satalecka, K. and Scalzotto, V. and Scapin, V. and Schultz, C. and Schweizer, T. and Shayduk, M. and Shore, S. N. and Sillanp{\"a}{\"a}, A. and Sitarek, J. and Snidaric, I. and Sobczynska, D. and Spanier, F. and Spiro, S. and Stamatescu, V. and Stamerra, A. and Steinke, B. and Storz, J. and Strah, N. and Sun, S. and Suri{\'c}, T. and Takalo, L. and Takami, H. and Tavecchio, F. and Temnikov, P. and Terzi{\'c}, T. and Tescaro, D. and Teshima, M. and Tibolla, O. and Torres, D. F. and Treves, A. and Uellenbeck, M. and Vankov, H. and Vogler, P. and Wagner, R. M. and Weitzel, Q. and Zabalza, V. and Zandanel, F. and Zanin, R. and Pfrommer, C. and Pinzke, A.},
  year = {2012},
  month = may,
  journal = {Astronomy \& Astrophysics},
  volume = {541},
  eprint = {1111.5544},
  primaryclass = {astro-ph},
  pages = {A99},
  issn = {0004-6361, 1432-0746},
  doi = {10.1051/0004-6361/201118502},
  urldate = {2023-12-19},
  archiveprefix = {arxiv},
  langid = {english},
  lccn = {2}
}

@article{magiccollaborationDeepObservationNGC2016,
  title = {Deep Observation of the {{NGC}} 1275 Region with {{MAGIC}}: Search of Diffuse Gamma-Ray Emission from Cosmic Rays in the {{Perseus}} Cluster},
  shorttitle = {Deep Observation of the {{NGC}} 1275 Region with {{MAGIC}}},
  author = {{MAGIC Collaboration} and Ahnen, M. L. and Ansoldi, S. and Antonelli, L. A. and Antoranz, P. and Babic, A. and Banerjee, B. and Bangale, P. and {de Almeida}, U. Barres and Barrio, J. A. and Gonz{\'a}lez, J. Becerra and Bednarek, W. and Bernardini, E. and Biasuzzi, B. and Biland, A. and Blanch, O. and Bonnefoy, S. and Bonnoli, G. and Borracci, F. and Bretz, T. and Buson, S. and Carmona, E. and Carosi, A. and Chatterjee, A. and Clavero, R. and Colin, P. and Colombo, E. and Contreras, J. L. and Cortina, J. and Covino, S. and Da Vela, P. and Dazzi, F. and De Angelis, A. and De Lotto, B. and Wilhelmi, E. de O{\~n}a and Mendez, C. Delgado and Di Pierro, F. and Dom{\'i}nguez, A. and Prester, D. Dominis and Dorner, D. and Doro, M. and Einecke, S. and Glawion, D. Eisenacher and Elsaesser, D. and {Fern{\'a}ndez-Barral}, A. and Fidalgo, D. and Fonseca, M. V. and Font, L. and Frantzen, K. and Fruck, C. and Galindo, D. and L{\'o}pez, R. J. Garc{\'i}a and Garczarczyk, M. and Terrats, D. Garrido and Gaug, M. and Giammaria, P. and Godinovi{\'c}, N. and Mu{\~n}oz, A. Gonz{\'a}lez and Gora, D. and Guberman, D. and Hadasch, D. and Hahn, A. and Hanabata, Y. and Hayashida, M. and Herrera, J. and Hose, J. and Hrupec, D. and Hughes, G. and Idec, W. and Kodani, K. and Konno, Y. and Kubo, H. and Kushida, J. and La Barbera, A. and Lelas, D. and Lindfors, E. and Lombardi, S. and Longo, F. and L{\'o}pez, M. and {L{\'o}pez-Coto}, R. and Lorenz, E. and Majumdar, P. and Makariev, M. and Mallot, K. and Maneva, G. and Manganaro, M. and Mannheim, K. and Maraschi, L. and Marcote, B. and Mariotti, M. and Mart{\'i}nez, M. and Mazin, D. and Menzel, U. and Miranda, J. M. and Mirzoyan, R. and Moralejo, A. and Moretti, E. and Nakajima, D. and Neustroev, V. and Niedzwiecki, A. and Rosillo, M. Nievas and Nilsson, K. and Nishijima, K. and Noda, K. and Orito, R. and Overkemping, A. and Paiano, S. and Palacio, J. and Palatiello, M. and Paneque, D. and Paoletti, R. and Paredes, J. M. and {Paredes-Fortuny}, X. and Pedaletti, G. and Persic, M. and Poutanen, J. and Moroni, P. G. Prada and Prandini, E. and Puljak, I. and Rhode, W. and Rib{\'o}, M. and Rico, J. and Garcia, J. Rodriguez and Saito, T. and Satalecka, K. and Schultz, C. and Schweizer, T. and Sillanp{\"a}{\"a}, A. and Sitarek, J. and Snidaric, I. and Sobczynska, D. and Stamerra, A. and Steinbring, T. and Strzys, M. and Takalo, L. and Takami, H. and Tavecchio, F. and Temnikov, P. and Terzi{\'c}, T. and Tescaro, D. and Teshima, M. and Thaele, J. and Torres, D. F. and Toyama, T. and Treves, A. and Acosta, M. Vazquez and Verguilov, V. and Vovk, I. and Ward, J. E. and Will, M. and Wu, M. H. and Zanin, R. and {and} and Pfrommer, C. and Pinzke, A. and Zandanel, F.},
  year = {2016},
  month = may,
  journal = {Astronomy \& Astrophysics},
  volume = {589},
  eprint = {1602.03099},
  primaryclass = {astro-ph},
  pages = {A33},
  issn = {0004-6361, 1432-0746},
  doi = {10.1051/0004-6361/201527846},
  urldate = {2023-09-09},
  archiveprefix = {arxiv},
  langid = {english},
  lccn = {2}
}

@article{themagiccollaborationMAGICGammarayTelescope2010,
  title = {{{MAGIC Gamma-ray Telescope Observation}} of the {{Perseus Cluster}} of {{Galaxies}}: Implications for Cosmic Rays, Dark Matter, and {{NGC1275}}},
  shorttitle = {{{MAGIC Gamma-ray Telescope Observation}} of the {{Perseus Cluster}} of {{Galaxies}}},
  author = {The {MAGIC Collaboration} and Aleksi{\'c}, J. and Antonelli, L. A. and Antoranz, P. and Backes, M. and Baixeras, C. and Balestra, S. and Barrio, J. A. and Bastieri, D. and Gonz{\'a}lez, J. Becerra and Bednarek, W. and Berdyugin, A. and Berger, K. and Bernardini, E. and Biland, A. and Bock, R. K. and Bonnoli, G. and Bordas, P. and Tridon, D. Borla and {Bosch-Ramon}, V. and Bose, D. and Braun, I. and Bretz, T. and Britzger, D. and Camara, M. and Carmona, E. and Carosi, A. and Colin, P. and Commichau, S. and Contreras, J. L. and Cortina, J. and Costado, M. T. and Covino, S. and Dazzi, F. and De Angelis, A. and {del Pozo}, E. De Cea and los Reyes, R. De and De Lotto, B. and De Maria, M. and De Sabata, F. and Mendez, C. Delgado and Doert, M. and Dom{\'i}nguez, A. and Prester, D. Dominis and Dorner, D. and Doro, M. and Elsaesser, D. and Errando, M. and Ferenc, D. and Fonseca, M. V. and Font, L. and Galante, N. and L{\'o}pez, R. J. Garc{\'i}a and Garczarczyk, M. and Gaug, M. and Godinovic, N. and Hadasch, D. and Herrero, A. and Hildebrand, D. and {H{\"o}hne-M{\"o}nch}, D. and Hose, J. and Hrupec, D. and Hsu, C. C. and Jogler, T. and Klepser, S. and Kr{\"a}henb{\"u}hl, T. and Kranich, D. and La Barbera, A. and Laille, A. and Leonardo, E. and Lindfors, E. and Lombardi, S. and Longo, F. and L{\'o}pez, M. and Lorenz, E. and Majumdar, P. and Maneva, G. and Mankuzhiyil, N. and Mannheim, K. and Maraschi, L. and Mariotti, M. and Mart{\'i}nez, M. and Mazin, D. and Meucci, M. and Miranda, J. M. and Mirzoyan, R. and Miyamoto, H. and Mold{\'o}n, J. and Moles, M. and Moralejo, A. and Nieto, D. and Nilsson, K. and Ninkovic, J. and Orito, R. and Oya, I. and Paiano, S. and Paoletti, R. and Paredes, J. M. and Partini, S. and Pasanen, M. and Pascoli, D. and Pauss, F. and Pegna, R. G. and {Perez-Torres}, M. A. and Persic, M. and Peruzzo, L. and Prada, F. and Prandini, E. and Puchades, N. and Puljak, I. and Reichardt, I. and Rhode, W. and Rib{\'o}, M. and Rico, J. and Rissi, M. and R{\"u}gamer, S. and Saggion, A. and Saito, T. Y. and Salvati, M. and {S{\'a}nchez-Conde}, M. and Satalecka, K. and Scalzotto, V. and Scapin, V. and Schultz, C. and Schweizer, T. and Shayduk, M. and Shore, S. N. and {Sierpowska-Bartosik}, A. and Sillanp{\"a}{\"a}, A. and Sitarek, J. and Sobczynska, D. and Spanier, F. and Spiro, S. and Stamerra, A. and Steinke, B. and Struebig, J. C. and Suric, T. and Takalo, L. and Tavecchio, F. and Temnikov, P. and Terzic, T. and Tescaro, D. and Teshima, M. and Torres, D. F. and Vankov, H. and Wagner, R. M. and Zabalza, V. and Zandanel, F. and Zanin, R. and Zapatero, J. and Pfrommer, C. and Pinzke, A. and En{\ss}lin, T. A. and Inoue, S. and Ghisellini, G.},
  year = {2010},
  month = feb,
  journal = {The Astrophysical Journal},
  volume = {710},
  number = {1},
  eprint = {0909.3267},
  primaryclass = {astro-ph},
  pages = {634--647},
  issn = {0004-637X, 1538-4357},
  doi = {2010012104130400},
  urldate = {2023-12-19},
  archiveprefix = {arxiv},
  langid = {english},
  lccn = {2}
}

@article{fleysherTestsStatisticalSignificance2004b,
  title = {Tests of {{Statistical Significance}} and {{Background Estimation}} in {{Gamma-Ray Air Shower Experiments}}},
  author = {Fleysher, R. and Fleysher, L. and Nemethy, P. and Mincer, A. I. and Haines, T. J.},
  year = {2004},
  month = mar,
  journal = {The Astrophysical Journal},
  volume = {603},
  number = {1},
  pages = {355},
  publisher = {IOP Publishing},
  issn = {0004-637X},
  doi = {10.1086/381384},
  urldate = {2024-04-27},
  langid = {english},
  lccn = {2}
}
\bibliographystyle{aasjournal}

\clearpage 

\appendix
\begin{appendices}
\section{Detailed Information OF THESE GALAXY CLUSTERS}
\label{apxA}

\renewcommand{\thefigure}{A.\arabic{figure}} 
\renewcommand{\thetable}{A.\arabic{table}}   
\setcounter{figure}{0} 
\setcounter{table}{0}  

In this appendix, we provide detailed information on the three galaxy clusters, as well as the distribution functions and parameters of $n_{e}$ and electron pressure $P_{e}$ used in this study. All basic information about these three clusters are listed in Table \ref{tab:info}.

To characterize the thermal gas within the ICM across different galaxy clusters, we employed various spatial models for $n_{e}$ and $P_{e}$, as used in previous studies \citep{thehesscollaborationConstraintsMultiTeVParticle2009, adamRayDetectionComa2021, consortiumProspectsGammaRay2023, collaborationConstrainingCosmicrayPressure2023c}. Details are provided in Table \ref{tab:distri}. For the Virgo cluster, previous studies focused only on the central region ($<$ 44 kpc). In this study, we adopt the latest fitting results for gas and temperature distributions, including data beyond the $R_{500}$ radius, obtained from Planck observations of the Sunyaev–Zel’dovich (SZ) effect \citep{adePlanckIntermediateResults2016}.

For the Perseus and Virgo clusters, the Table \ref{tab:distri} shows the distribution of electron thermal temperature. The pressure distribution can be derived from $n_{e}$ and $T$ using the following equation:
\begin{equation}
\centering
    P_e(r)=n_e(r) k_{\mathrm{B}} T(r)
\end{equation}

\begin{table*}[!htb]
\centering
\caption {The basic properties of Coma, Perseus and Virgo cluster.}
\begin{tabular}{cccccccc}
\toprule[0.5mm]
Name    & RA    & DEC   & Redshift & Size            &  Angular size      & Mass                &$E_{th_{500}}^{a}$ \\
        & (deg) & (deg) & z    & $R_{500}$(Mpc)  &$\theta_{500}$(deg) & $M_{500}$($M_{\odot}$)  &  (erg)   \\
\hline 
Coma    &194.95 &27.98  & 0.023    &   1.35          &  0.78 & $6.13 \times 10^{14}$ & $2.69 \times 10^{63}$ \\ 
Perseus &49.95  &41.51  & 0.017    &   1.26          &  0.96 & $5.77 \times 10^{14}$ & $2.08 \times 10^{63}$ \\
Virgo   &187.70 &12.39  & 0.004    &   0.66          &  2.3  & $0.83 \times 10^{14}$ & $8.42 \times 10^{61}$ \\
\bottomrule[0.4mm]
\end{tabular}
\\
{\footnotesize 
$^a$ Total thermal energy within $R_{500}$ radius.\\
}
\label{tab:info}
\end{table*}

\begin{table*}[!htb]
\centering
\caption{The spatial distribution models and corresponding parameters for the three galaxy clusters used in this study.}
\begin{threeparttable}
\begin{tabular}{llll}
\toprule[0.5mm]
\textbf{Cluster} & \textbf{Physical quantity} & \textbf{Function} & \textbf{Parameters} \\
\toprule[0.3mm]
\multirow{2}{*}{Coma\tnote{a}} 

& Density profile: & $n_{e}(r) = n_0 \left[ 1 + \left( \frac{r}{r_c} \right)^2 \right]^{-\frac{3\beta}{2}}$ & $n_{0}=3.42 \times 10^{-3} \mathrm{~cm}^{-3}, r_{\mathrm{c}}=290\ \mathrm{kpc}, \beta=0.75$ \\
& Perssure profile: & $n_{e}(r) = \frac{P_0}{\left( \frac{r}{r_p} \right)^c \left( 1 + \left( \frac{r}{r_{p}} \right)^a \right)^{\frac{b-c}{a}}}$ & \parbox[t]{6cm}{ $P_{0} = 0.022\ \mathrm{kev}\ \mathrm{cm}^{-3}, r_{p} = 466.8\ \mathrm{kpc}, a = 1.8, b = 3.1, c = 0.0$} \\ 

\hline
\multirow{2}{*}{Perseus\tnote{b}}
& Density profile: & \parbox[t]{6cm}{$n_{e}(r) = n_{0,1} \left[ 1 + \left( \frac{r}{r_{c,1}} \right)^2 \right]^{-\frac{3\beta_1}{2}} \\ \quad\quad\quad + n_{0,2} \left[ 1 + \left( \frac{r}{r_{c,2}} \right)^2 \right]^{-\frac{3\beta_2}{2}}$} & \parbox[t]{6cm}{$n_{0,1}=4.6\times10^{-2}\ \mathrm{cm}^{-3}, r_{c,1}=57\ \mathrm{kpc}, \beta_{1}=1.2, n_{0,2}=3.6 \times 10^{-3}\ \mathrm{cm}^{-3}, r_{c,2}=278\ \mathrm{kpc}, \beta_{2} = 0.71$} \\


& Temperature profile: & \parbox[t]{6cm}{$k_{\mathrm{B}} T(r)=7 \times \left(1+\left(\frac{r}{r_{t, 1}}\right)^3\right) \\ \times \left(2.3+\left(\frac{r}{r_{t, 1}}\right)^3\right)^{-1}\left(1+\left(\frac{r}{r_{t, 2}}\right)^{1.7}\right)^{-1} \mathrm{keV}$} & $r_{t,1} = 73.8\  \mathrm{kpc}, r_{t,2} = 1600\ \mathrm{kpc}$ \\ 

\hline

\multirow{2}{*}{Virgo\tnote{c}}
& Density profile: & \parbox[t]{6cm}{ $n_{e}(r) = n_{0,1} \left[ 1 + \left( \frac{r}{r_{c,1}} \right)^2 \right]^{-\frac{3\beta_1}{2}} \\ \quad\quad\quad + n_{0,2} \left[ 1 + \left( \frac{r}{r_{c,2}} \right)^2 \right]^{-\frac{3\beta_2}{2}} (r\lesssim30\ kpc)
\\ n_{\mathrm{e}}(r)=\frac{8.5 \times 10^{-5}}{(r / \mathrm{Mpc})^{1.2}} \mathrm{~cm}^{-3} (r\gtrsim250\ kpc)$ } & \parbox[t]{6cm}{$n_{0,1}=8.9\times10^{-2}\ \mathrm{cm}^{-3}, r_{c,1}=4\ \mathrm{kpc}, \beta_{1}=1, n_{0,2}=1.9 \times 10^{-3}\ \mathrm{cm}^{-3}, r_{c,2}=19.4\ \mathrm{kpc}, \beta_{2} = 0.47$} \\ 

& Temperature profile: & $k_{\mathrm{B}} T(r)=\frac{2.4-0.77 \times \mathrm{e}^{-(r / \mathrm{Mpc})^2 /\left(2 \sigma_r^2\right)}}{1+[0.9 \times(r / \mathrm{Mpc})]^2} \mathrm{keV}$ &  $\sigma_{r}$ = 23.7 kpc. \\ 

\bottomrule[0.4mm]
\end{tabular}
\begin{tablenotes}
    \item[a]  For Coma, We adopt the $\beta$ model \citep{cavaliereDistributionHotGas1978a} for the density profile and a generalized Navarro Frenk White \citep{nagaiTestingXRayMeasurements2007} model for pressure radial profile, with parameters taken from \cite{brielObservationComaCluster1992} and \cite{planckcollaboration:PlanckIntermediateResults2013a}.
    \item[b]  The functions and parameters used for the Perseus cluster in this study are consistent with those employed in \cite{consortiumProspectsGammaRay2023}.
    \item[c] \cite{adePlanckIntermediateResults2016} combines a double-beta function for the inner region $(r\lesssim30\ kpc)$ with a power-law function for the outer region $(r\gtrsim250\ kpc)$ to describe the thermal electron number density distribution in the Virgo cluster. A simple linear interpolation was applied to the intermediate region. We converted the angular units into distance units.
\end{tablenotes}
\end{threeparttable}
\label{tab:distri}
\end{table*}

\section{EBL model}
\label{apxB}

\renewcommand{\thefigure}{B.\arabic{figure}} 
\renewcommand{\thetable}{B.\arabic{table}}   
\setcounter{figure}{0} 
\setcounter{table}{0}  

\begin{figure*}[h!]
    \centering
     \includegraphics[width=0.5\linewidth]{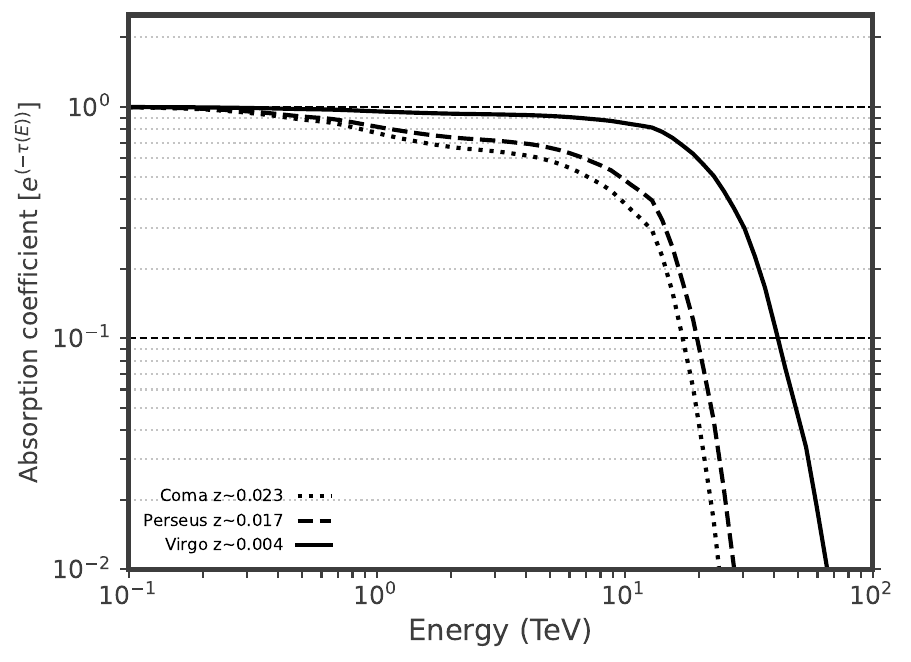}
    \caption{The EBL absorption efficiency for each of the three galaxy clusters. Coma is represented by dotted line, Perseus by dashed line, and Virgo by solid line.}
    \label{fig:ebl}
\end{figure*}

In this appendix, we provide detailed information regarding the EBL model. \grays within the energy range observed by LHAASO undergo attenuation due to pair-production interactions with the CMB and EBL as they traverse extragalactic space. The photon-photon pair-production cross-section averaged over directions of the background-radiation field depends on the product of energies of colliding photons. For the given energy of the $\gamma$-ray photon $E_\gamma$, it peaks at the  wavelength of background photons $\lambda \sim 1 (E_{\gamma}/{\rm 1 \ TeV})^{-1} \mu \rm m$.
Thus,  the $\gamma$-ray opacity above and below 100 TeV is due to the absorption on CMB and EBL, respectively.  The  opacity has been calculated by performing the line-of-sight integral of the product of the pair production cross section with the energy density of the radiation fields. The dependence of the pair production cross section on the energy is given in \cite{gouldPairProductionPhotonPhoton1967a}. The EBL energy density is taken from the \citet{franceschiniExtragalacticOpticalinfraredBackground2008}. Figure \ref{fig:ebl} illustrates the variation in absorption efficiency with energy at the respective distances of the Coma, Perseus, and Virgo clusters, as predicted by the EBL model adopted in this paper.

\end{appendices}



\end{document}